\documentclass[aps,prd,preprintnumbers,superscriptaddress,nofootinbib,notitlepage,floatfix]{revtex4-2}
\usepackage[pdftex]{graphicx}
\usepackage{bm,latexsym,amsmath,amssymb,amsfonts,mathrsfs}
\usepackage{color}
\usepackage{empheq}
\allowdisplaybreaks[1]
\usepackage[pdftex,colorlinks=true,linkcolor=blue,citecolor=cyan,backref=page]{hyperref}
\newcommand*{\D}{\mathrm{d}}
\newcommand*{\mpl}{M_{\mathrm{Pl}}}
\newcommand*{\ga}{g_{A}}
\newcommand{\LL}{\left}
\newcommand{\RR}{\right}
\newcommand\figref[1]{Fig.~\ref{#1}}

\newcommand*{\nmi}{R_i}
\newcommand*{\nm}{R}
\newcommand*{\nma}{\bar{R}}
\newcommand*{\trf}{\beta}
\newcommand*{\nmAi}{{\cal R}_i}
\newcommand*{\nmA}{{\cal R}}

\begin{document}
\title{How does SU(\textit{N})-natural inflation isotropize the universe?
}
%
\author{Tomoaki~Murata}
\email[Email: ]{tmurata@rikkyo.ac.jp}
\affiliation{Department of Physics, Rikkyo University, Toshima, Tokyo 171-8501, Japan
}
\author{Tomohiro~Fujita}
\email[Email: ]{tomofuji@aoni.waseda.jp}
\affiliation{Waseda Institute for Advanced Study, Waseda University, Shinjuku, Tokyo 169-8050, Japan
}
\affiliation{Research Center for the Early universe, University of Tokyo, Bunkyo, Tokyo 113-0033, Japan
}
\author{Tsutomu~Kobayashi}
\email[Email: ]{tsutomu@rikkyo.ac.jp}
\affiliation{Department of Physics, Rikkyo University, Toshima, Tokyo 171-8501, Japan
}
%
\begin{abstract}
We study the homogeneous and anisotropic dynamics of pseudoscalar inflation coupled to an SU($N$) gauge field.
To see how the initially anisotropic universe is isotropized in such an inflation model,
we derive the equations to obtain axisymmetric SU($N$) gauge field configurations
in Bianchi type-I geometry
and discuss a method to identify their isotropic subsets
which are the candidates of their late-time attractor.
Each isotropic solution is characterized by the corresponding SU(2) subalgebra of the SU($N$) algebra.
It is shown numerically that the isotropic universe is a universal late-time attractor
in the case of the SU(3) gauge field.
Interestingly, we find that a transition between the two distinct gauge-field configurations
characterized by different SU(2) subalgebras can occur during inflation.
We clarify the conditions for this to occur.
This transition could leave an observable imprint on the CMB
and the primordial gravitational wave background.
\end{abstract}
\preprint{RUP-22-23}
\maketitle
\section{Introduction}

It is widely believed that
inflation~\cite{Starobinsky:1980te,Guth:1980zm,Sato:1980yn},
a prolonged phase of the quasi-de Sitter expansion,
occurred in the early universe.
The inflationary mechanism solves the problems concerning the
initial conditions of the standard Big Bang model, i.e.,
the flatness and horizon problems, and thus
naturally explains the homogeneity and isotropy of the universe.
Inflation can also generate primordial density fluctuations
that are consistent with observations such as the CMB anisotropies.

For a given inflationary model to be successful,
it is required that no particular fine-tuned initial conditions
are imposed at the beginning of inflation. For example,
inflation would be spoiled if initial anisotropies were not diluted away
in the early universe. Fortunately, according to Wald's cosmic no-hair
theorem~\cite{Wald:1983ky},
all Bianchi models except type IX isotropize toward de Sitter
in the presence of a positive cosmological constant.
However, one has to be careful when applying this theorem to
models of inflation, because the actual inflationary expansion
is caused not by a cosmological constant but by a potential energy
of a dynamical scalar field.
In the case of power-law inflation, this point was addressed in Refs.~\cite{Moss:1986ud,Kitada:1991ih,Rendall:2004ic}.
The inflationary isotropization process is particularly nontrivial
in the presence of vector fields.
In fact, it is possible to construct an inflationary model
with a U(1) gauge field coupled to a scalar field
through a particular coupling function
that violates the cosmic no-hair theorem, though
the amount of remaining anisotropies is slow-roll suppressed~\cite{Watanabe:2009ct}
(see Ref.~\cite{Yamamoto:2012tq} for the
extension of multiple U(1) gauge fields).
This anisotropic inflationary model has been extended
to the cases of a SU(2) gauge field~\cite{Maeda:2013daa} 
and a SU(3) gauge field~\cite{Gao:2021qwl} coupled to 
a scalar field.
These examples show that an inflationary universe does not
always isotropize in the presence of a vector field~\cite{Maleknejad:2011jr,Maleknejad:2012as}.

The chromo-natural inflation (CNI) model~\cite{Adshead:2012kp}
is a model of inflation well motivated by particle physics in which a pseudo-scalar axion and an SU(2) gauge field are
coupled through the Chern-Simons coupling
(see Ref.~\cite{Maleknejad:2012fw} for a review).
It has been shown in Refs.~\cite{Maleknejad:2013npa,Wolfson:2020fqz,Wolfson:2021fya}
that the Bianchi type-I metric in the CNI model generically isotropize
and the isotropic solution is an attractor in favor of the cosmic no-hair theorem.
Although the CNI model is an interesting class of inflationary models,
the model was already excluded according to observations~\cite{Adshead:2013nka}
(see, however, some extended models compatible with observations~\cite{Dimastrogiovanni:2016fuu,Adshead:2016omu}).

The CNI model has been extended to the case with
a general SU($N$) gauge field in Ref.~\cite{Fujita:2021eue}.
It has been shown that this SU($N$)-natural inflation model
admits an isotropic configuration of the gauge field
forming an SU(2) subalgebra in SU($N$)
on a homogeneous and isotropic spacetime~\cite{Fujita:2021eue}.
The amplitude and energy density of the gauge field
depend on the gauge coupling $g_A$ and a parameter
$\lambda$, which characterizes the embedding of an SU(2) subalgebra,
through the combination of $g_A\lambda$.
Linear perturbations of SU($N$)-natural inflation
have been investigated in Ref.~\cite{Fujita:2022fff}, showing that
the perturbation dynamics and linear predictions depend on $\lambda$
only through the same combination $g_A\lambda$.
As the dynamics of SU($N$)-natural inflation is thus obtained simply
by replacing $g_A$ of CNI with $g_A\lambda$,
the former model cannot be distinguished observationally from the latter
unless, for instance, a gauge-field configuration of one of the SU(2) subalgebras
is unstable and transits to another isotropic configuration
during inflation.

In this paper, we study the homogeneous dynamics of SU($N$)-natural inflation
in more detail, improving the previous analysis of~\cite{Fujita:2021eue}
in two respects.
First, we investigate whether or not
the isotropic solution is the late-time attractor when starting from
anisotropic initial configurations. Our analysis is more rigorous than
that of Ref.~\cite{Fujita:2021eue} because
we solve the Einstein equations and the equations of motion for the gauge fields and
the scalar field consistently.
Second, we clarify the conditions for the transition from
an isotropic gauge-field configuration of one of the SU(2) subalgebras
to another to occur.

This paper is organized as follows. 
In the next section, we review the isotropic solution of the
SU($N$)-natural inflation model.
We then derive the conditions for a configuration of SU($N$) gauge fields
to be axisymmetric, and consider the isotropic limit
as a candidate for the late-time attractor
in Sec.~\ref{sec:axisymmetric_SU(3)}.
We also show that the spacetime metric is istropized once the $x$, $y$, and $z$-components of the gauge fields have the same norm.
In Sec.~\ref{sec:axisymmetric_SU(N)}, 
by restricting ourselves to the case of SU(3),
we provide several examples of axisymmetric gauge field configurations.
In Sec.~\ref{sec:num_cal}, we
present the results of our numerical analysis on the
dynamics of the axisymmetric SU(3)-natural inflation.
On the basis of our numerical results,
we draw our conclusions in Sec.~\ref{sec:conclusions}.

\section{Review of SU(\textit{N})-natural inflation}\label{sec:su(n)-natural}

In this section, we review the SU($N$)-natural inflation model and introduce 
its homogeneous and isotropic cosmological solution~\cite{Fujita:2021eue}.
The SU($N$)-natural inflation model is described by the Lagrangian
\begin{align}
  {\cal L} = -\frac{1}{4} F_{\mu \nu}^{a} F_a^{\mu \nu}
  -\frac{1}{2} \LL(\partial\phi\RR)^2 -V(\phi)
  -\frac{\phi}{4 f} F_{\mu \nu}^{a} \widetilde{F}^{\mu \nu}_a,
\end{align}
where $\phi$ is the inflaton scalar field, $V(\phi)$ is its potential,
and $f$ is a constant parameter.
The field strength $F_{\mu \nu}^{a}$ of the SU($N$) gauge field $A_\mu^a$
and its dual $\widetilde{F}^{\mu \nu}_{a}$ are defined by
\begin{align}
  F_{\mu \nu}^{a} &=\partial_{\mu} A_{\nu}^{a}-\partial_{\nu} A_{\mu}^{a}
  +\ga f^{abc} A_{\mu}^{b} A_{\nu}^{c},
  \\
  \widetilde{F}^{\mu \nu}_{a} &= \frac{1}{2} \varepsilon^{\mu \nu \rho \lambda}F_{\rho \lambda}^{a},
\end{align}
where $f^{abc}$ is the structure constant of the SU($N$) algebra,
$\ga$ is the gauge coupling constant, and
$\varepsilon^{\mu\nu\rho\lambda}=\epsilon^{\mu\nu\rho\lambda}/\sqrt{-g}$
is the Levi-Civit\`{a} tensor with $\epsilon^{0123}=1$.

Let us consider homogeneous and isotropic cosmology in this model. The Friedmann‐Lema\^{i}tre‐Robertson-Walker metric is given by
\begin{align}
  \D s^2 = -\D t^2 +a^2(t) \left(\D x^2 + \D y^2 +\D z^2\right),
  \label{met:FLRW}
\end{align}
where $a(t)$ is the scale factor.
Here we assume the spatial flatness. It is nontrivial to include the spatial curvature in the universe with a similar gauge-field configuration~\cite{Murata:2021vnb}.
Throughout the paper, we use the temporal gauge, $A_0^a=0$.
It has been shown in Ref.~\cite{Fujita:2021eue}
that the SU(2) subalgebra of the SU($N$) algebra allows for an isotropic solution.
The isotropic gauge field configuration can be expressed as
\begin{align}
  A_{0}(t)=0, 
  \quad 
  A_{i}(t) = a(t) \psi (t) {\cal T}_{i},
  \label{eq:iso_ansatz}
\end{align}
where ${\cal T}_{i}$ is the generator of the SU(2) subalgebra.
The SU(2) generators can be expressed as a linear combination of
the SU($N$) generators $T^a$: ${\cal T}_{i} = n^a_i T^a$.
The SU(2) generators satisfy the commutation relation and the normalization condition,
\begin{align}
  \LL[{\cal T}_i,{\cal T}_j\RR]
  = i \lambda \epsilon_{ijk}{\cal T}_k,
  \quad
  \operatorname{Tr}\LL({\cal T}_i {\cal T}_j\RR)
  = \frac{1}{2}\delta_{ij},
  \label{eq: T property}
\end{align}
where $\lambda$
is a number that is
different depending on the choice of the SU(2) subalgebra
of SU($N$).

Using the ansatz~\eqref{eq:iso_ansatz},  we obtain
the equations of motion for $\phi$ and $\psi$ as
\begin{align}
  &\ddot{\phi} + 3H\dot{\phi} + V_{,\phi} 
  = -\frac{3\ga\lambda}{f} \psi^2 \LL(\dot{\psi} + H\psi\RR),
  \label{eq:axion_eom_iso}
  \\
  &\ddot{\psi} + 3H\dot{\psi} + \LL(2H^2 +\dot{H}\RR)\psi + 2\ga^2\lambda^2\psi^3
  = \frac{\ga\lambda}{f} \dot{\phi} \psi^2,
\end{align}
where the dot represents differentiation with respect to $t$
and $H:=\dot a/a$ is the Hubble parameter.
The same equations can be obtained just by replacing $g_A$ with $g_A\lambda$
in the inflaton and gauge-field equations of motion in
original SU(2) chromo-natural inflation~\cite{Adshead:2012kp}.
Therefore, one would expect qualitatively the same dynamics of
homogeneous and isotropic cosmology as that in the previous chromo-natural model.
Assuming slow roll, one obtains a solution for the gauge field
staying at the minimum of its effective potential~\cite{Adshead:2012kp},
\begin{align}
  \psi_{\rm min}^3 = -\frac{f V_{,\phi}}{3\ga\lambda H}.
  \label{eq:psi_min}
\end{align}
The solution can also be derived by dropping all the time derivatives
in Eq.~\eqref{eq:axion_eom_iso}.

To simplify the following analysis,
we use the linear potential,
\begin{align}
  V(\phi) = -\frac{\mu^4}{f}\phi,
  \label{eq:axion_pot}
\end{align}
where $\mu$ is a constant having the dimension of mass.
In this simplified setup,
we would expect that
the isotropic solution for the gauge field
in the slow-roll approximation reduces to
\begin{align}
  H^{1/3}\psi_{\rm min} = \LL(\frac{\mu^4}{3\ga\lambda}\RR)^{1/3}
  =\textrm{const}.
  \label{eq:iso_const}
\end{align}
Since $\mu$ and $g_A$ are constant model parameters, the slow-roll value of $H^{1/3}\psi$ depends only on $\lambda$ as $H^{1/3}\psi_{\rm min}\propto \lambda^{-1/3}$.

Having introduced the SU($N$)-natural inflation model and reviewed the isotropic solution, let us move to discuss axisymmetric cosmological solutions in that model.

\section{Axisymmetric configurations of SU(\textit{N}) gauge fields}
\label{sec:axisymmetric_SU(3)}

One of the purposes of this paper is to discuss the robustness of
the isotropic solution in SU($N$)-natural inflation to initial anisotropies.
This point was studied in Ref.~\cite{Fujita:2021eue}
for randomly oriented initial gauge field configurations
with the {\em fixed} scalar-field evolution on a {\em fixed}
de Sitter background. In this paper, we improve the previous analysis
by consistently solving all the relevant equations, i.e., the Einstein equations and
the equations of motion for the gauge fields and the inflaton.
To handle this generalization, we focus on an axisymmetric setup.
In this section, we rigorously derive the equations to obtain
axisymmetric SU($N$) gauge field configurations and discuss a method to
identify their isotropic subsets which are the candidates of their late-time attractor.

\subsection{Conditions for axisymmetric SU(\textit{N}) gauge fields}\label{sec:deriv_axisymm_su(3)}

Following Refs.~\cite{Murata:2011wv,Gao:2021qwl}, 
let us consider axisymmetric gauge-field configurations
in the axisymmetric Bianchi type-I geometry,
\begin{align}
  \D s^2 = -\D t^2 +a^2(t)
  \left[e^{-4\sigma (t)}\D x^2 +e^{2\sigma (t)}
  \left(\D y^2 +\D z^2\right)\right],
  \label{met:bianchi type-I}
\end{align}
where $\sigma(t)$ characterizes the deviation from the isotropic space.
Note that the metric~\eqref{met:bianchi type-I} is isotropic when
$\dot\sigma=0$, because constant $\sigma$ can always be absorbed
into a rescaling of the spatial coordinates.

Since we consider homogeneous SU($3$) gauge fields in the temporal gauge ($A_0^a=0$),
the gauge fields can be written as
\begin{align}
  A = A_x(t)\D x + A_y(t)\D y + A_z(t)\D z.
  \label{eq:all comp. of gauge fields}
\end{align}
We assume that the
principal axes of metric and gauge-field anisotropies point to the same direction.
An infinitesimal rotational transformation around this axis is generated by 
the Killing vector $\xi = -z\partial_y + y\partial_z$.
Under this rotational transformation, the gauge field transforms as
\begin{align}
  A^a \mapsto A^a+\epsilon \mathscr{L}_\xi A^a,
  \qquad
  \mathscr{L}_\xi A^a 
  = A_z^a \D y -A_y^a \D z,
  \label{eq:Lie deriv. of gauge field}
\end{align}
where $\epsilon$ is an infinitesimal parameter.
Even if the gauge field configuration is not invariant under the above transformation, the configuration is regarded to be axisymmetric as long as there exists a gauge transformation that cancels the effect of the rotational transformation.
There is a residual global gauge symmetry, $A\to UAU^\dagger$,
and in the case of an infinitesimal transformation
with $U=1+i \epsilon u^a T^a$ the gauge field transforms as
\begin{align}
   A^a \mapsto A^a+\epsilon \delta _g A^a,
  \qquad
  \delta _g A^a = -f^{abc} u^b A^c,
  \label{eq:global gauge tr.}
\end{align}
where $u^a$ is a constant vector.
To maintain the rotational invariance,
we require that the two transformations are compensated:
\begin{align}
  \mathscr{L}_\xi A^a + \delta _g A^a =0  .
  \label{eq:rot to gauge}
\end{align}
We thus obtain the conditions for the axisymmetric configuration as
\begin{align}
  f^{abc}u^b A_x^{c} = 0,
  \quad
  f^{abc}u^b A_y^{c} = A_z^a,
  \quad
  f^{abc}u^b A_z^{c} = -A_y^a.
  \label{eq:fuA_eqs}
\end{align}
For given $u^a$, the above equations can be used to determine the possible configurations of $A$.
We see that
$A_x^a$ is determined independently.
Eliminating $A_z^a$, we get the equation solely for $A_y^a$:
\begin{align}
  A_y^a =  M^{ab} A_y^{b},
  \quad
  M^{ab} := f^{ace} f^{bde} u^c u^d.
  \label{eq:eigenvalue eq.}
\end{align}
The explicit form of the
$(N^2-1)\times (N^2-1)$ matrix $M^{ab}$ depends on $u^a$.
This equation may be considered as an eigenvalue equation, and then $A_y^a$ must be an eigenvector with eigenvalue one.
We shall solve the above equations explicitly
in the case of SU(3) gauge fields in the next section.

\subsection{Conditions for isotropic subsets}\label{sec:iso_conf}

In the previous subsection, we derived the conditions for
axisymmetric configurations of the SU($N$) gauge field.
In this subsection, we
further discuss conditions for the isotropic configurations,
which are included in the axisymmetric SU($N$) configurations as subsets.
In the isotropic solution~\eqref{eq:iso_ansatz}, the gauge field takes the form of $A_i=\sqrt{2}\nmA\mathcal{T}_i$ with the common norm $\sqrt{2}\nmA := a\psi$
and the generators $\mathcal{T}_i$ of the SU(2) subalgebra.
Multiplying the commutation relation~\eqref{eq: T property}
of the SU(2) generators by the squared norm $2\nmA^2$,
we find the condition satisfied by the isotropic solution,
\begin{align}
    \LL[A_i,A_j\RR] = \sqrt{2}i\lambda \epsilon_{ijk}
     \nmA A_k.
    \label{eq:comm_su(2)}
\end{align}
This condition enables us to remove anisotropic components and identify the isotropic configuration in the axisymmetric configurations.
In the next section, we will use Eq.~\eqref{eq:comm_su(2)} in 
some cases of axisymmetric SU(3) gauge field configurations
to identify their isotropic subsets.

\subsection{Isotropization of the metric}
\label{Isotropization of metric}

We now show that the spatial metric is isotropized when the norms of the gauge field are equal in the following sense.
In an axisymmetric Bianchi type-I universe, we decompose the gauge field in the temporal gauge $(A^a_0=0)$ as
\begin{align}
  A_i^a \D x^i
  = \psi_x^a (t) \LL[a(t) e^{-2\sigma (t)} \RR] \D x
  + \psi_y^a (t) \LL[a(t) e^{\sigma(t)} \RR] \D y
  + \psi_z^a (t) \LL[a(t) e^{\sigma(t)} \RR] \D z,
  \label{eq:su(n)_conf_gen}
\end{align}
where $i=1,2,3$ and $a = 1,\dots ,N^2-1$.
Note that $\psi_i^a$ does not include the metric variables (i.e. $a$ and $\sigma$).
The above equation can be understood as a redefinition of the gauge field $A_i^a$ into $\psi_i^a$, and its benefit will be seen below.
We define the norm of the gauge field for each $i$ as
\begin{align}
  \nmi(t)
  := \left[ \frac{1}{2}\sum_{a=1}^{N^2-1}\LL(\psi_i^a\RR)^2 \right]^{1/2}
  \quad(\text{no sum over } i).\label{def:norm-a}
\end{align}
This expression implies that $\sqrt{2}\nmi$ corresponds to the radial coordinate in the $\LL(N^2-1\RR)$-dimensional spherical coordinate system when $\psi_i^a$ is viewed as the Cartesian coordinate.
One can thus parametrize the components of
the SU($N$) gauge field
using the spherical coordinates for each $i$ as
\begin{align}
  \psi_i^a=\sqrt{2}\nmi(t)n_i^a(t),\qquad 
  n_i^a=\left\{
  \begin{array}{ll}
    \displaystyle
    \cos \theta^{a}_{i}
    \prod_{b=1}^{a-1} \sin \theta^{b}_{i} & (a=1, \dots, N^2-2)
    \\
    \displaystyle
    \prod_{b=1}^{N^2-2} \sin \theta^{b}_{i} & (a=N^2-1)
  \end{array}
  \right.
  \quad(\text{no sum over } i),
  \label{eq:polar_coord_comp2}
\end{align}
where $0\leq \theta^a_{i}(t) \leq \pi$ ($a=1,\dots,N^2-3$) and
$0\leq \theta^{N^2-2}_{i}(t) < 2\pi$.
Furthermore, it is convenient to decompose $\nmi$ into
the geometric mean and the anisotropic part as
\begin{align}
    \nmi=\nma(t)e^{\trf_i(t)},
    \quad 
    \nma:=\left[R_{x}R_{y}R_{z}\right]^{1/3},
    \quad 
    \sum_i^3 \trf_i=0.
\end{align}
From the symmetry, we assume that $R_{y}=R_{z}$.
To summarize, we can express the component of the gauge field as
\begin{align}
  A^a_i = \psi^a_i a e^{\Sigma_i}
  = \sqrt{2} a\nma e^{\trf_i + \Sigma_i} n_i^a
  \quad(\text{no sum over } i),
\end{align}
where we introduced the vector $\Sigma_i = \LL(-2\sigma , \sigma, \sigma\RR)$
that represents the anisotropic part of the spatial metric.
Thanks to the redefinition~\eqref{eq:su(n)_conf_gen}, we now easily extract the terms involving the anisotropic part of the spatial metric $\sigma$
from our Lagrangian to get
\begin{align}
  {\cal L}_{\sigma} 
  &= 3\mpl^2 \dot\sigma^2
  - 2\dot\sigma 
  \partial_t \LL(R_{x}^2 - R_{y}^2\RR)
  -4H \dot\sigma \LL(R_{x}^2 - R_{y}^2\RR)
  + \dot\sigma^2 \LL(4R_{x}^2 + 2R_{y}^2\RR)
  .
\end{align}
Since there is a symmetry under a constant shift
of the anisotropic part of the spatial metric, $\sigma\to\sigma+\,$const,
we have a conserved quantity (denoted as ${\cal D}$).
One can thus integrate the equation of motion for $\sigma$
to obtain
\begin{align}
  &\dot\sigma = \frac{{\cal D}/a^3 
  +\partial_t\LL(R_{x}^2 - R_{y}^2\RR)
  +2H \LL(R_{x}^2 -R_{y}^2\RR)}
  {3\mpl^2 + 4R_{x}^2 + 2R_{y}^2},
\end{align}
If the anisotropic part of the gauge field decays away,
$\trf_{i}\to 0$, then one has $R_{i}\to \bar{R}$,
and hence $\dot\sigma\to 0$ at late times.
The spatial metric is therefore isotropized
if the norm of the gauge field defined by Eq.~\eqref{def:norm-a}
settles down to $R_{x}=R_{y}=R_{z}$ regardless of $n_i^a$.
We can extend this result to the general Bianchi type-I case (see Appendix~\ref{app:c}).

In Sec.~\ref{sec:iso_conf}, we discussed the condition for the isotropic solution forming the SU(2) subgroup. That condition not only requires that the three norms are equal $R_i=\bar{R}$ but also determines the internal configuration of the gauge field $n^a_i$ such that $\mathcal{T}_i=n_i^a T^a$.
However, we have seen above that requiring only the former condition leads to the isotropic space.
Therefore, the condition of the SU(2) subgroup~\eqref{eq:comm_su(2)} is in fact a sufficient condition for the metric isotropization.

Here we stress that $\sigma$ can be disregarded in some cases.
One can always find a coordinate system in which $\sigma(t)=0$ once $\sigma$ stops evolving, since we have the freedom to shift the value of $\sigma$ by a coordinate rescaling.
In this convenient coordinate system, we can easily study some conditions for the isotropic gauge field in the isotropized universe without having $\sigma$.
In what follows, therefore, when we discuss the isotropic configuration of the gauge field, we work in such a coordinate system where $\sigma$ vanishes, given that the universe has been isotropized. In such cases, we simply have $A_i^a=a\psi_i^a$.
In contrast, when we discuss axisymmetric configurations of the gauge field, we cannot eliminate $\sigma$
and will take it into account.
This is because axisymmetric configurations will be used as an initial condition of our numerical calculations in which the universe is still anisotropic, and then $A_i^a$ contains nonzero $\sigma$ as in Eq.~\eqref{eq:su(n)_conf_gen}.

\section{Examples of axisymmetric configurations of SU(3) gauge field  and their isoropic subsets}
\label{sec:axisymmetric_SU(N)}

In this section,
we compute axisymmetric configurations and their isotropic subsets of
the SU(3) gauge field by solving the conditions derived in Sec.~\ref{sec:axisymmetric_SU(3)}.
In principle, by solving Eq.~\eqref{eq:eigenvalue eq.} for generic $u^a$
one can find the 
general axisymmetric configuration of the SU($N$) gauge field. 
However, it is difficult to analytically find the general solution to Eq.~\eqref{eq:eigenvalue eq.}. For this reason, we consider the SU(3) case with
some simple choices of $u^a$.
See Appendix~\ref{app:su3} for a brief summary of the SU(3) algebra.

In the next section, we will perform numerical calculations to clarify whether and how the universe is isotropized in SU(3)-natural inflation by using axisymmetric configurations as the initial conditions.
It is known that the SU(2) subgroups in SU(3) are classified into two types characterized respectively by $\lambda=1$ and $\lambda=1/2$.
The axisymmetric configurations desirable for our purpose contain the isotropic solutions with both $\lambda=1$ and $1/2$ as the subsets, and we therefore seek such configurations that have a parameter allowing us to switch between the two possible types. Fortunately, we can find such
a case with a simple choice of $u^a$, as we will see below.

\subsection{Example 1: $u^1=2$ and others$\,=0$}\label{sec:iso_u1=2}

Let us consider the case
where $u^1$ is nonvanishing, but the other $u^a$'s are zero,
namely,
\begin{equation}
u^a=(k,0,0,0,0,0,0,0),\label{uk000}
\end{equation}
with $k$ being a real number.
Then, the matrix $M$ appearing in Eq.~\eqref{eq:eigenvalue eq.} 
has only its diagonal components as
\begin{align}
  M^{ab} = {\rm diag.}\LL(0,k^2,k^2,
  \frac{k^2}{4},\frac{k^2}{4},\frac{k^2}{4},\frac{k^2}{4},0\RR).
  \label{eq: Mab}
\end{align}
As discussed below Eq.~\eqref{eq:eigenvalue eq.}, the solution $A_y^a$ of Eq.~\eqref{eq:eigenvalue eq.} may be regarded
as an eigenvector whose eigenvalue is unity.
Therefore, we have two branches of the solutions: $k=\pm 1$ and $k=\pm 2$.
Note that the cases with a minus sign only invert the sign of $A_z$,
and hence it suffices to focus on the cases with a plus sign.

It is pedagogical to describe the calculation for $u^1=k=2$ in detail. In this case, 
since the fourth, fifth, sixth, and seventh diagonal components of the matric $M$ are unity, 
only $A_y^4, A_y^5, A_y^6, A_y^7$ are nonzero and the other components of $A_y$ vanish.
Then, $A_x$ and $A_z$ can be easily obtained by solving Eq.~\eqref{eq:fuA_eqs}. One finds the axisymmetric configuration of the SU(3) gauge field for $u^1=k=2$ as
\begin{align}
  A_x = A_x^1T^1 + A_x^8T^8,
 \quad
  A_y = A_y^4T^4 + A_y^5T^5 + A_y^6T^6 + A_y^7T^7,
 \quad 
  A_z = -A_y^7T^4 + A_y^6T^5 - A_y^5T^6 + A_y^4T^7,
  \label{eq:conf_u1_k=2}
\end{align}
where the six coefficients, $A_x^1, A_x^8, A_y^4, A_y^5, A_y^6, A_y^7$, are arbitrary.
We have thus obtained an example of the axisymmetric configurations of the SU(3) gauge field.

To determine the isotropic solutions within the axisymmetric configuration~\eqref{eq:conf_u1_k=2},
we substitute Eq.~\eqref{eq:conf_u1_k=2} into the commutation relation~\eqref{eq:comm_su(2)}. Two branches of the solution are found. One branch has $A_x^8\neq 0$ and the other has $A_x^8=0$:
\begin{empheq}[left={\empheqlbrace}]{alignat=2}
  & \lambda^2\nmA^2=2\LL(A_x^1\RR)^2 = \LL(A_y^4\RR)^2 + \LL(A_y^5\RR)^2,
  \quad
  A_x^8 = \pm \sqrt{3} A_x^1,
  \quad
  A_y^6 = \pm A_y^4,
  \quad
  A_y^7 = \pm A_y^5, &\qquad &(A_x^8\neq 0) 
  \label{eq:iso_con_A8=0-u}
  \\
  & \lambda^2\nmA^2=\frac{1}{8}\LL(A_x^1\RR)^2
  = \frac{1}{8}\LL[\LL(A_y^4\RR)^2 + \LL(A_y^5\RR)^2
  +\LL(A_y^6\RR)^2 + \LL(A_y^7\RR)^2\RR],\quad
  A_y^4 A_y^6 + A_y^5 A_y^7=0,&       &(A_x^8= 0)
  \label{eq:iso_con_A8=0-d}
\end{empheq}
where the common norm of $A_i$ is given by
\begin{equation}
    \nmA^2 =
    \nmAi^2 = \frac{1}{2}\sum_{a=1}^{8}\LL(A_i^a\RR)^2
      \quad
      (\text{no sum over } i).
\end{equation}
Using this expression for $\nmA$, one finds that Eqs.~\eqref{eq:iso_con_A8=0-u}
and~\eqref{eq:iso_con_A8=0-d} correspond to $\lambda=1$ and $\lambda= 1/2$, respectively.

Specializing the axisymmetric configuration~\eqref{eq:conf_u1_k=2}
to the case of Eq.~\eqref{eq:iso_con_A8=0-u} ($\lambda=1$),
we obtain 
\begin{align}
  &A_x = A_x^1 \big(T^1 + \sqrt{3}T^8 \big),
  \\
  &A_y = A_y^4\LL(T^4 + T^6\RR) + A_y^5\LL(T^5 + T^7\RR),
  \label{eq:conf_u1_k=2_iso}\\
  &A_z = -A_y^5\LL(T^4 + T^6\RR)+ A_y^4\LL(T^5 + T^7\RR),
\end{align}
To rewrite the above expression in the manifestly isotropic form, $A_i=\sqrt{2}\nmA\mathcal{T}_i$, we define the SU(2) generators
${\cal T}_i$ as
\begin{align}
  {\cal T}_1 &= \frac{T^1 + \sqrt{3}T^8}{2},
  \\
  {\cal T}_2 &= \cos{\theta} \, \frac{T^4 + T^6}{\sqrt{2}}
  + \sin{\theta} \, \frac{T^5 + T^7}{\sqrt{2}},
  \\
  {\cal T}_3 &= -\sin{\theta} \, \frac{T^4 + T^6}{\sqrt{2}}
  + \cos{\theta} \, \frac{T^5 + T^7}{\sqrt{2}},
\end{align}
where we introduced spherical coordinates $A_y^4 = \nmA \cos{\theta}, A_y^5 = \nmA\sin{\theta}$ with $\nmA=\sqrt{2}A_x^1$.
It is easy to see that ${\cal T}_i$ satisfies the properties of the SU(2) generators, Eq.~\eqref{eq: T property}, with $\lambda=1$.

We repeat the same procedure in the case of
Eq.~\eqref{eq:iso_con_A8=0-d} ($\lambda=1/2$) and find that
the gauge field can be written in the form
$A_i=\sqrt{2}\nmA\mathcal{T}_i$ with
\begin{align}
  A_x^1 = \sqrt{2}\nmA,
  \quad
  A_y^4 = A_x^1 \cos{\theta} \cos{\varphi},
  \quad
  A_y^5 = A_x^1 \cos{\theta} \sin{\varphi},
  \quad
  A_y^6 = -A_x^1 \sin{\theta} \sin{\varphi},
  \quad
  A_y^7 = A_x^1 \sin{\theta} \cos{\varphi}.
  \label{eq:iso_con_ctr}
\end{align}
The SU(2) generators here are given by
\begin{align}
  &{\cal T}_1 = T^1,
  \\
  &{\cal T}_2 =
  \cos{\theta} \cos{\varphi} \, T^4 
  + \cos{\theta} \sin{\varphi} \, T^5
  - \sin{\theta} \sin{\varphi} \, T^6
  + \sin{\theta} \cos{\varphi} \, T^7,
  \\
  &{\cal T}_3 =
  - \sin{\theta} \cos{\varphi} \, T^4
  - \sin{\theta} \sin{\varphi} \, T^5
  - \cos{\theta} \sin{\varphi} \, T^6 
  + \cos{\theta} \cos{\varphi} \, T^7.
\end{align}
These generators satisfy the SU(2) subalgebra~\eqref{eq: T property} with 
$\lambda=1/2$.

As we have seen, the present example admits the two SU(2) subgroups, which is exactly what we want.

\subsection{Example 2: $u^1=1$ and others$\,=0$}\label{sec:iso_u1=1}

Let us consider again the case where $u^a$ is
of the form of Eq.~\eqref{uk000} and
hence the matrix is given by Eq.~\eqref{eq: Mab},
but now $k=1$.
The calculation can be done in a similar way to the case with $k=2$.
The axisymmetric configuration for $u^1=k=1$ reads
\begin{align}
  A_x = A_x^1T^1 + A_x^8T^8, 
  \quad
  A_y = A_y^2T^2 + A_y^3T^3, 
  \quad 
  A_z = -A_y^3T^2 + A_y^2T^3.\label{eq:conf_u1_k=1}
\end{align}

Having thus obtained the axisymmetric configuration,
we further impose the conditions~\eqref{eq:comm_su(2)} for the isotropic subsets, 
we obtain 
\begin{align}
  &\LL(A_x^1\RR)^2 = \LL(A_y^2\RR)^2 +\LL(A_y^3\RR)^2,
  \quad
  A_x^8 = 0,
  \label{eq:com_rel_u1=1}
\end{align}
and then
the isotropic solution of the form $A_i=\sqrt{2}\nmA {\cal T}_i$
can be obtained with the SU(2) generators
\begin{align}
  {\cal T}_1 = T^1,
  \quad
  {\cal T}_2 = \cos{\theta} \, T^2 + \sin{\theta} \, T^3,
  \quad
  {\cal T}_3 = -\sin{\theta} \, T^2 + \cos{\theta} \, T^3.
\label{eq:com_rel_u1=1-2}
\end{align}
These SU(2) generators satisfy the commutation relation~\eqref{eq: T property} with $\lambda=1$, and no other isotropic solution is available in the
$u^1=1$ case. Thus, the axisymmetric configuration for $u^1=1$ does not include the isotropic solution with $\lambda=1/2$.

\subsection{Comments on the other cases}

In the previous two subsections, we consider
$u^a$ with the single nonvanishing component, $u^1\neq 0$.
Let us comment on the other cases.
In the case where only a single component of $u^a$
other than $u^1$ and $u^8$ is nonvanishing,
we have essentially the same results
as in the above two examples.
Therefore, for $u^a=2$ ($a=2,\dots,7$) there is an axisymmetric
configuration that involves both $\lambda=1$ and $\lambda=1/2$
isotropic solutions.
In contrast,
in the case where only $u^8$ is nonvanishing,
we have an isotropic configuration forming only
the $\lambda=1$ SU(2) subalgebra.
One could also consider more general cases
where multiple components of $u^a$ are nonvanishing.
There are some (rather simpler) cases
that are analytically tractable,
but in general the analysis would be so complicated
that an exhaustive investigation is hindered.
It should be emphasized, however, that 
we have already obtained the physically interesting
axisymmetric configuration that admits
both $\lambda=1$ and $\lambda=1/2$ isotropic limits
in the case of $u^1=2$.

\section{Numerical results}\label{sec:num_cal}

In the previous section, we obtained the axisymmetric configurations and the associated isotropic configurations of the
SU(3) gauge field. We are interested in whether or not the SU(3) gauge field in an axisymmetric configuration at some initial moment evolves into the isotropic solution associated with that.
To address this question, we run numerical calculations.
Interestingly, there exists a nontrivial variety of the ways how the SU(3) gauge field isotropizes the universe, as we will see below.

The parameters used in our numerical calculations are listed in Table~\ref{tab:parameters}.
In the following 
numerical calculations we use the units in which $\mpl =1.$
First, we consider the case with $u^1=1$ discussed in Sec.~\ref{sec:iso_u1=1}.
We rewrite the gauge field as
\begin{align}
  A_x^1 = ae^{-2\sigma}\psi_1, \quad
  A_y^2 = ae^{\sigma}\psi_2, \quad
  A_y^3 = ae^{\sigma}\psi_3, \quad
  A_x^8 = ae^{-2\sigma}\psi_8.
  \label{eq:psi1238}
\end{align}
Note that the energy density of the gauge field remains
constant if
$\psi_a=\,$const $(a=1,2,3,8)$ as well as $H=\,$const and $\dot{\sigma}=\,$const.
We solve the equation of motion
for the inflaton $\phi$, gauge fields $\psi_a$,
and the metric variables $a$ and $\sigma$,
which are given by Eqs.~\eqref{eq:phi_u1_k=1}--\eqref{eq:sigma_u1_k=1}
in Appendix~\ref{sec:eom_u1=1}.
The initial condition for the scalar field at $t=0$ is
set as $(\phi (0) ,\dot{\phi} (0))=(-3\mpl ,0)$,
and the velocity of the gauge field is set as
$\dot{\psi}_a(0)=0$, so that the gauge constraints Eq.~\eqref{eq:gauge_constraint_u1=1} are satisfied.
We set the initial amplitudes of the gauge fields so that
all of them are of the same order.
The scale factor at $t=0$ is set as $a(0)=1$.
The initial conditions are thus placed in the slow-roll regime,
with the energy density of the gauge field being much smaller than that of the scalar field.
We therefore choose to use
the initial value of the Hubble parameter
that is of the same order as its slow-roll value,
$H_i:=H(0)=\sqrt{3}\mu^2/\sqrt{f} = 10^{-6}\mpl$.
We then determine the initial metric anisotropy $\dot\sigma(0)$
by using the Friedmann equation~\eqref{eq:Friedmann_u1=1}.
From these initial conditions and Eq.~\eqref{eq:metric_aniso_u1=1},
the integration constant $\mathcal{D}$ is determined.
In our numerical calculations,
we confirm that the Friedmann equation
and Eq.~\eqref{eq:metric_aniso_u1=1} are satisfied
with sufficiently high accuracy at each time step.

\begin{table}[htb]
  \centering
    \caption{Parameters}
  \label{tab:parameters}
  \begin{tabular}{|c|c|c|} \hline
    $f$ & $\mu$ & $g_A$  \\ \hline
    $3\times 10^{-4}\mpl$ & $10^{-4}\mpl$ & $2\times 10^{-6}$\\ \hline
  \end{tabular}
\end{table}

\begin{figure}[tb]
    \begin{center}
        \includegraphics[keepaspectratio=true,height=45mm]{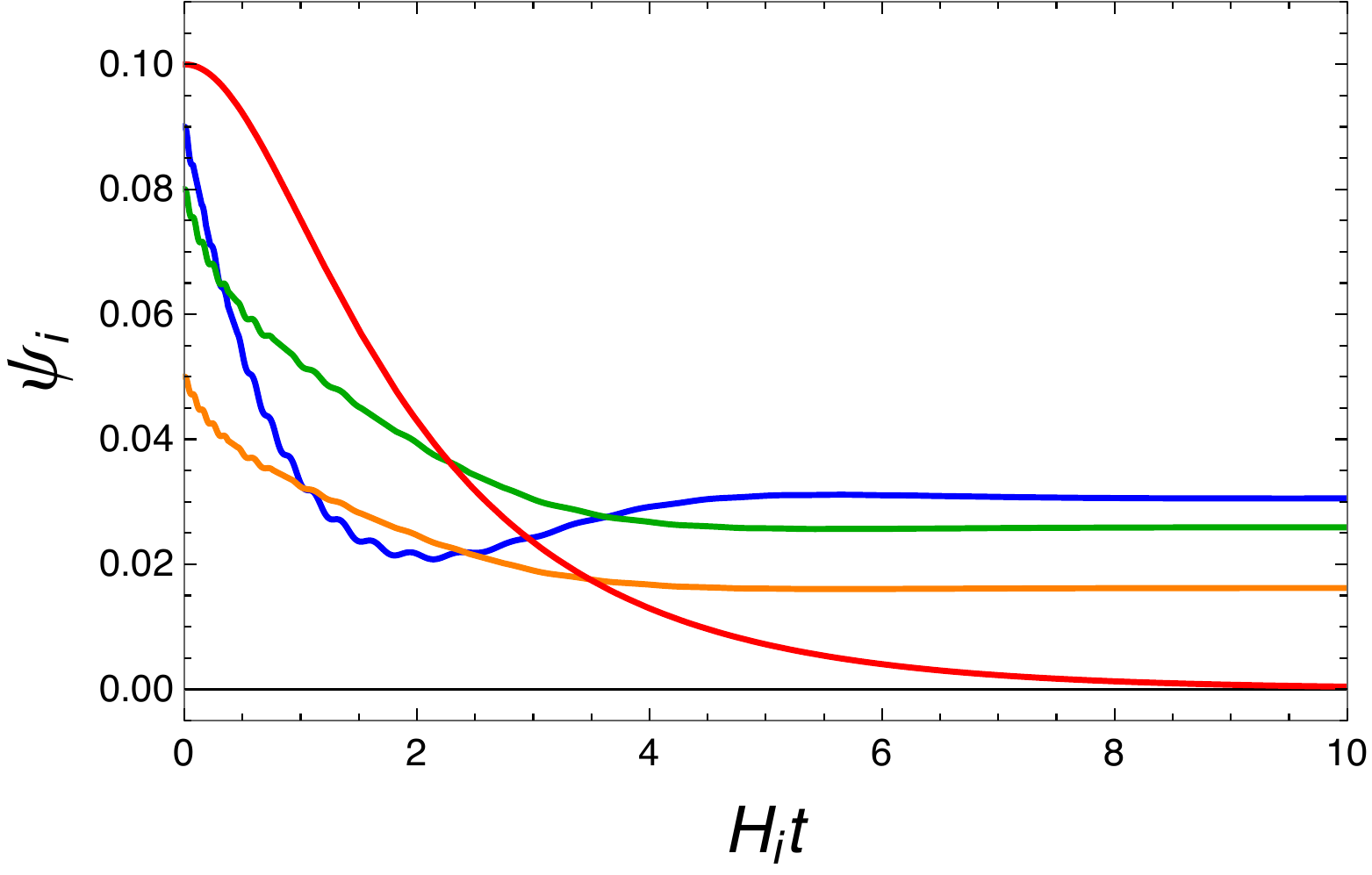}
        \includegraphics[keepaspectratio=true,height=45mm]{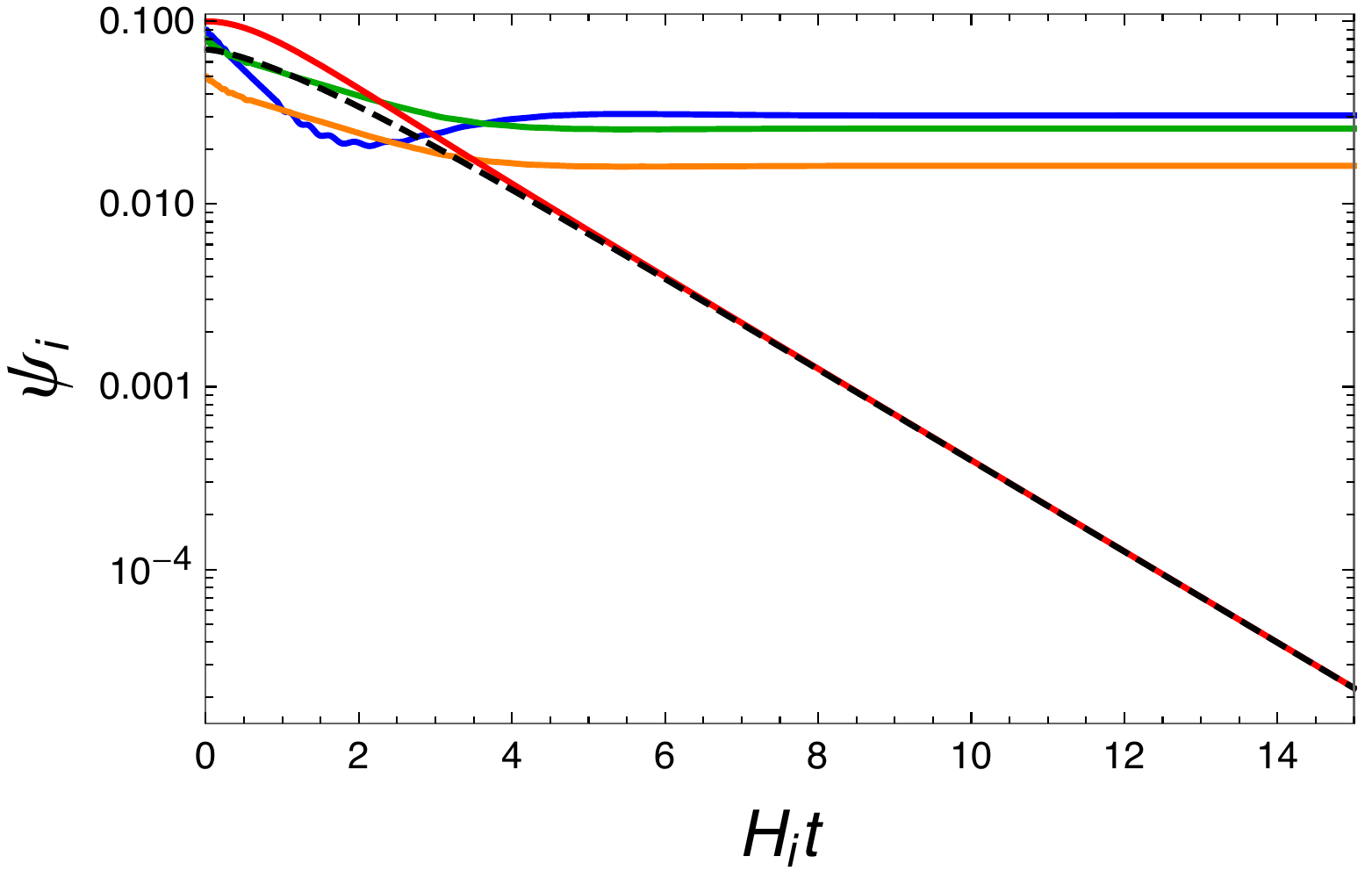}
    \end{center}
     \caption{
     The evolution of the gauge field amplitudes,
     $\psi_1$ (blue), $\psi_2$ (orange), 
     $\psi_3$ (green), and $\psi_8$ (red) defined in Eq.~\eqref{eq:psi1238},
     in linear scale (left) and in log scale (right).
     The black dashed line shows the late-time behavior $\psi_8\propto 1/a$ [Eq.~\eqref{eq:p8_latetime}].
     The initial condition is the $u^1=1$ axisymmetric configuration~\eqref{eq:conf_u1_k=1} with its four components being of the same order. After some transient evolution for a few e-folds, $\psi_1, \psi_2$, and $\psi_3$ stay constant, forming the isotropic solution. However, $\psi_8$, which is not compatible with the SU(2) subalgebra for this initial configuration, quickly decays.}  
      \label{fig:plot12num_u1_k=1_amp.pdf}
 \end{figure}

\begin{figure}[tb]
    \begin{center}
        \includegraphics[keepaspectratio=true,height=45mm]{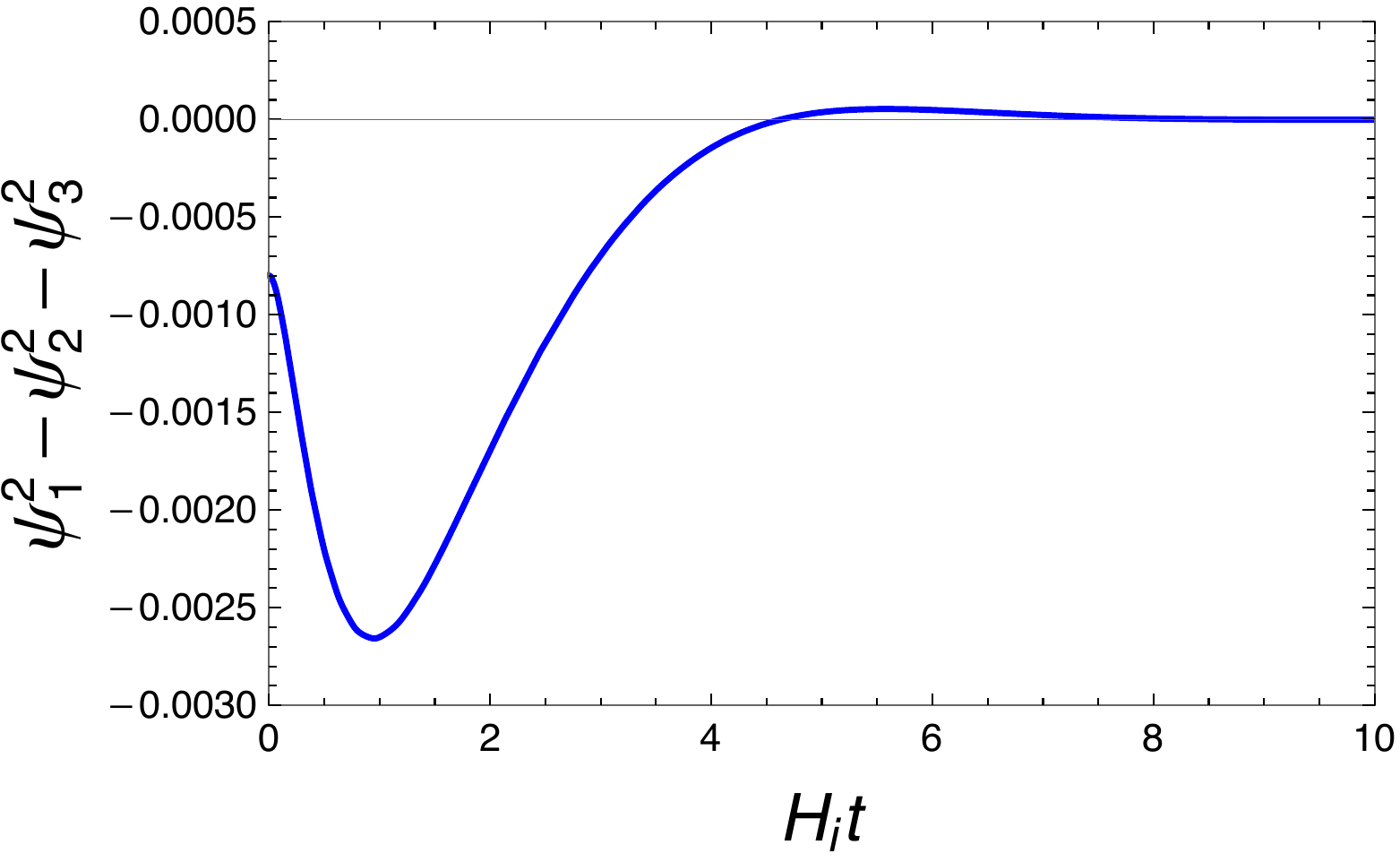}
        \includegraphics[keepaspectratio=true,height=45mm]{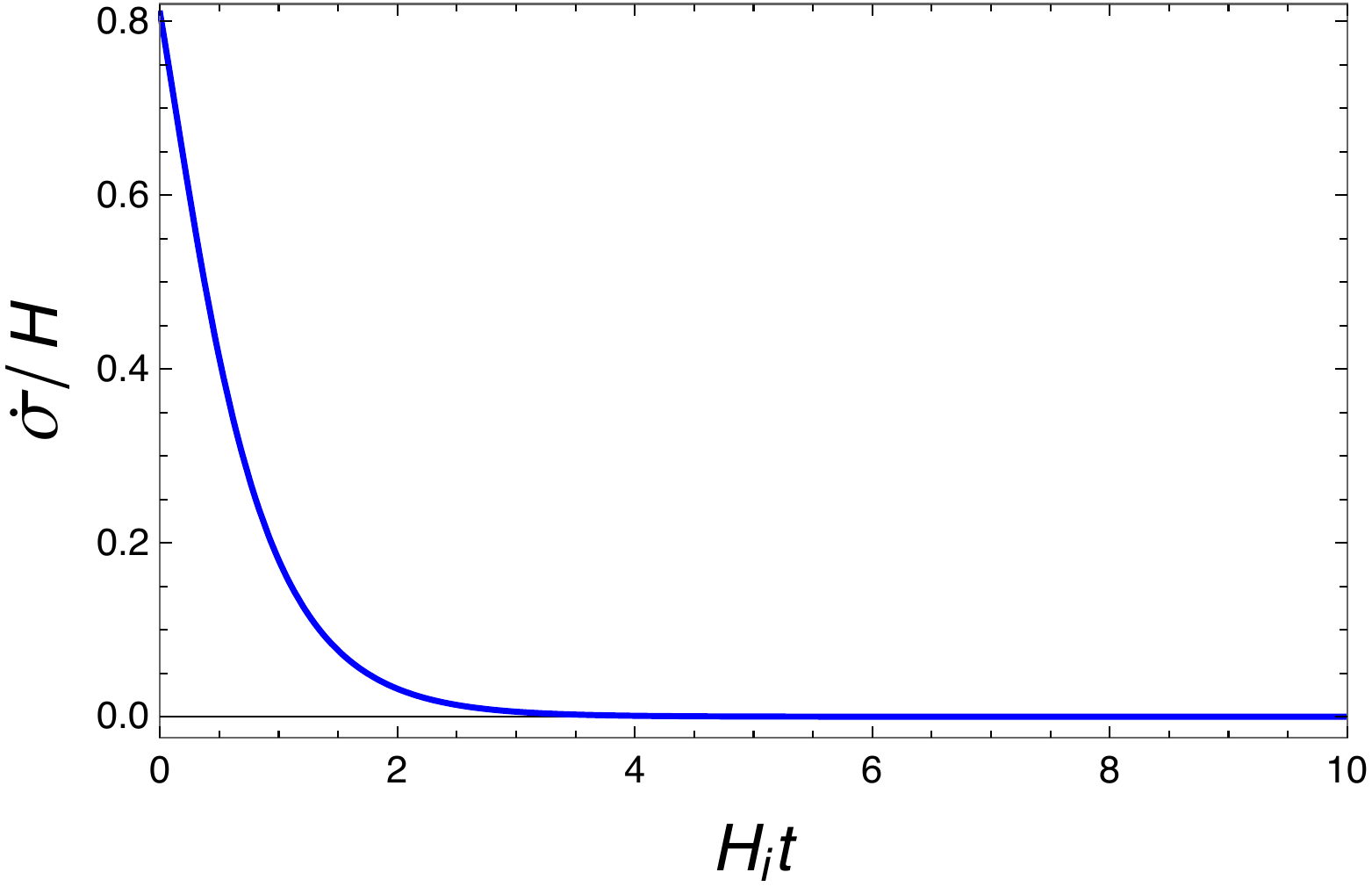}
    \end{center}
      \caption{
      The left panel shows $\psi_1^2-\psi_2^2-\psi_3^2$.
      The right panel presents the evolution of the metric anisotropy $\dot\sigma$.
      These results indicate the universe is isotropized in a few e-folds.
      The initial condition is the same as that taken in \figref{fig:plot12num_u1_k=1_amp.pdf}.
      }
      \label{fig:plot12num_u1_k=1_ds.pdf}
  \end{figure}

In \figref{fig:plot12num_u1_k=1_amp.pdf}, we present an example of our numerical results for the time evolution of the gauge field in terms of $\psi_1$, $\psi_2$, $\psi_3$, and $\psi_8$. 
The initial configuration is given by Eq.~\eqref{eq:conf_u1_k=1} with $(\psi_1,\psi_2,\psi_3,\psi_8)=(0.09\mpl , 0.08\mpl , 0.05\mpl , 0.1\mpl)$.
One can see from \figref{fig:plot12num_u1_k=1_amp.pdf} that $\psi_1$, $\psi_2$, and $\psi_3$
settle down to constant values within several e-foldings,
while $\psi_8$ continues to decay.
Indeed, we can analytically argue that $A_x^8$ should decay in this case of the initial configuration~\eqref{eq:conf_u1_k=1}.
The Chern-Simons term reads
\begin{align}
  F^a_{\mu\nu}\widetilde{F}_a^{\mu\nu}
  &=\frac{2}{3}\frac{\ga}{\sqrt{-g}} \partial_t\left[\epsilon^{ijk}
  f_{abc} A^a_i A^b_j A^c_k \right].
\end{align}
Since the structure constants of SU(3) satisfy $f_{81*}=f_{82*}=f_{83*}=0$,
$A_x^8$ does not appear in this term or in the self-coupling terms between the gauge fields [see Eq.~\eqref{psi8 EoM u1=1 case})].
Therefore, $A_x^8$ cannot be sourced 
and is diluted with the cosmic expansion.
It is easy to see analytically the late-time behavior of $\psi_8$.
Ignoring $\dot{H}$, $\ddot{\sigma}$, and $\dot{\sigma}$ at late times,
the solution of Eq.~\eqref{psi8 EoM u1=1 case} is given by 
\begin{align}
  \psi_8 \propto \frac{1}{a}, 
  \label{eq:p8_latetime}
\end{align}
where we discarded another independent solution
that decays more quickly.
One can see from the right panel of Fig.~\ref{fig:plot12num_u1_k=1_amp.pdf}
that our numerical solution reproduces this behavior at late times.
In the left panel of \figref{fig:plot12num_u1_k=1_ds.pdf}, 
we show the difference between the norm of $A_x$ and that of $A_y$
in terms of
$2(\nm_{x}^2-\nm_{y}^2)=\psi_1^2-(\psi_2^2+\psi_3^2)$,
from which we see that $\nm_{x}^2-\nm_{y}^2\to 0$
within several e-foldings.
In the right panel of \figref{fig:plot12num_u1_k=1_ds.pdf},
we plot the time evolution of the metric anisotropy $\dot\sigma$,
which is seen to decrease to zero, indicating that the metric is isotropized.
We have confirmed that the same behavior can be seen in all our numerical runs with different initial values of the gauge field.
We thus conclude that this initially anisotropic configuration
is attracted to the isotropic solution.
The result matches the associated isotropic configuration with the SU(2) subalgebra predicted in Eq.~\eqref{eq:com_rel_u1=1-2}.
Therefore, we confirm that
the axisymmetric configuration of the SU(3) gauge field evolves into the isotropic one forming the SU(2) subgroup.\footnote{Note that our numerical solution at late times does not satisfy $(A_x^1)^2-(A_y^2)^2-(A_z^3)^2=0$ [Eq.~\eqref{eq:com_rel_u1=1}] in general.
This is because, while the spatial part of the isotropic metric is assumed to be of the form $a^2(\D x^2+\D y^2+\D z^2)$ when writing Eq.~\eqref{eq:com_rel_u1=1}, in the actual numerical calculation we have $a^2[e^{-4\sigma}\D x^2+e^{2\sigma}(\D y^2+\D z^2)]$ with $\sigma=\,$const at late times. After rescaling the spatial coordinates appropriately, our numerical solution does satisfy Eq.~\eqref{eq:com_rel_u1=1}.
}

Next, let us consider the case with $u^1=2$.
We solve the equations of motion~\eqref{eq:phi_u1_k=2}--\eqref{eq:sigma_u1_k=2} given in Appendix~\ref{sec:eom_u1=2}.
The parameters and the initial conditions are the same as those in
the numerical calculations for the $u^1=1$ case except for the initial amplitudes of the gauge fields as explained below.
It is convenient to introduce the normalized mean amplitude of the gauge field as
\begin{align}
  \Psi(t) := \LL(\frac{3\ga H(t)}{\mu^4}\RR)^{1/3}
  \sqrt{\sum_{a,i}\frac{\LL[\psi_i^a(t)\RR]^2}{3}}.
  \label{eq:mean_amplitude}
\end{align}
The normalization factor here
is chosen so that we expect $\Psi\to \lambda^{-1/3}$
as the solution approaches to either of the isotropic ones
[see Eq.~\eqref{eq:iso_const}]. This quantity is therefore useful
for discriminating between the isotropic solutions with different $\lambda$.
The initial conditions range over $\Psi(0) \in [0.5,5]$ with positive random amplitudes of the gauge field components.

\begin{figure}[tb]
    \begin{center}
        \includegraphics[keepaspectratio=true,height=58mm]{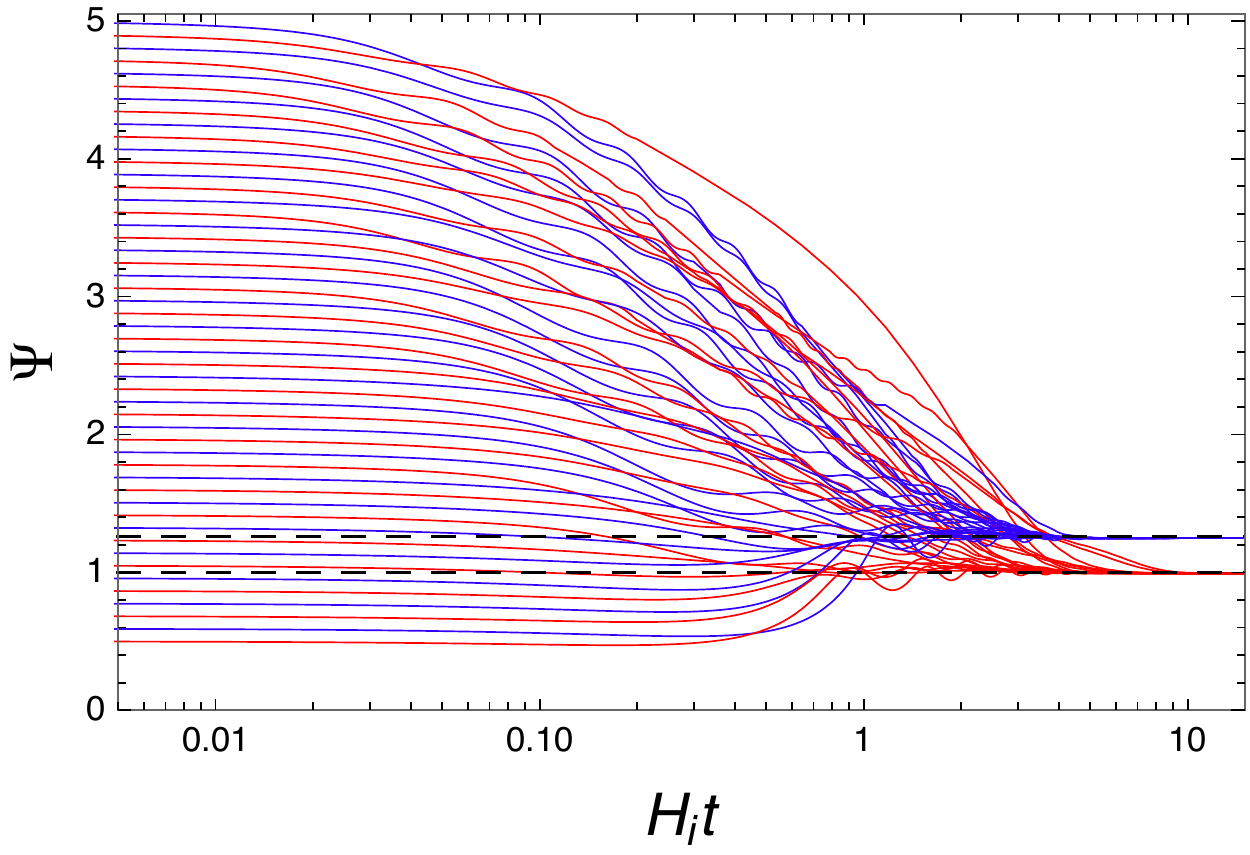}
    \end{center}
     \caption{
     The evolution of the mean amplitude $\Psi(t)$ of the gauge field defined in Eq.~\eqref{eq:mean_amplitude}.
     The initial conditions of 
     $\Psi$ are chosen in the range $\Psi(0) \in [0.5,5]$ with positive random amplitudes of the gauge field.
     The additional initial condition
     $\psi_5=\psi_6=\psi_8=0$ is imposed only for the blue lines.
     The black dashed horizontal lines represent the isotropic solutions in Eq.~\eqref{eq:iso_const}; the upper and lower line shows the $\lambda=1/2$ and $\lambda=1$ solutions, respectively.
     All the cases with $\psi_5=\psi_6=\psi_8=0$ (blue lines) converge into the $\lambda=1/2$ solution, while the $\lambda=1$ solution is the attractor for generic initial conditions.
 	}
      \label{fig:plot32num_u1_compare_init.pdf}
 \end{figure}

In \figref{fig:plot32num_u1_compare_init.pdf}, we display 50 realizations of the mean amplitude of the gauge field~\eqref{eq:mean_amplitude}.
Each line represents the time evolution of $\Psi(t)$ and the colors denote the initial values of the gauge field:
the solutions with the initial conditions $A_y^5=A_y^6=A_x^8=0$ are colored in blue and
the solutions for which
at least one of $A_y^5$, $A_y^6$, and $A_x^8$ is nonvanishing initially
are colored in red.
Aside from these restrictions, we randomly distribute the initial values of the gauge field components that are included in the $u^1=2$ axisymmetric configuration~\eqref{eq:conf_u1_k=2}.
We can see that all the
lines converge into either of the two isotropic attractor solutions inferred in Sec.~\ref{sec:iso_u1=2}.
The branching of the solutions is determined by the initial conditions of the gauge fields.
We find that the solution with $\lambda =1/2$ is realized
when setting $A_y^5=A_y^6=A_x^8=0$ at the initial moment.
The condition $A_y^5=A_y^6=A_x^8=0$ then continues to hold for $t>0$. Indeed,
by inspecting Eqs.~\eqref{eq:A5_u1_k=2},~\eqref{eq:A6_u1_k=2}, and~\eqref{eq:A8_u1_k=2}
in Appendix~\ref{sec:eom_u1=2},
we see that $\psi_5(t)=\psi_6(t)=\psi_8(t)=0$ is the solution of the system with
the initial conditions $\psi_5=\psi_6=\psi_8=\dot\psi_5=\dot\psi_6=\dot\psi_8=0$.
This solution satisfies Eq.~\eqref{eq:iso_con_A8=0-d}, and hence in the isotropic limit the system settles down to the $\lambda=1/2$ solution.
Our numerical calculations imply that other generic initial conditions result in
the $\lambda=1$ isotropic solution.

\begin{figure}[tb]
    \begin{center}
        \includegraphics[keepaspectratio=true,height=37mm]{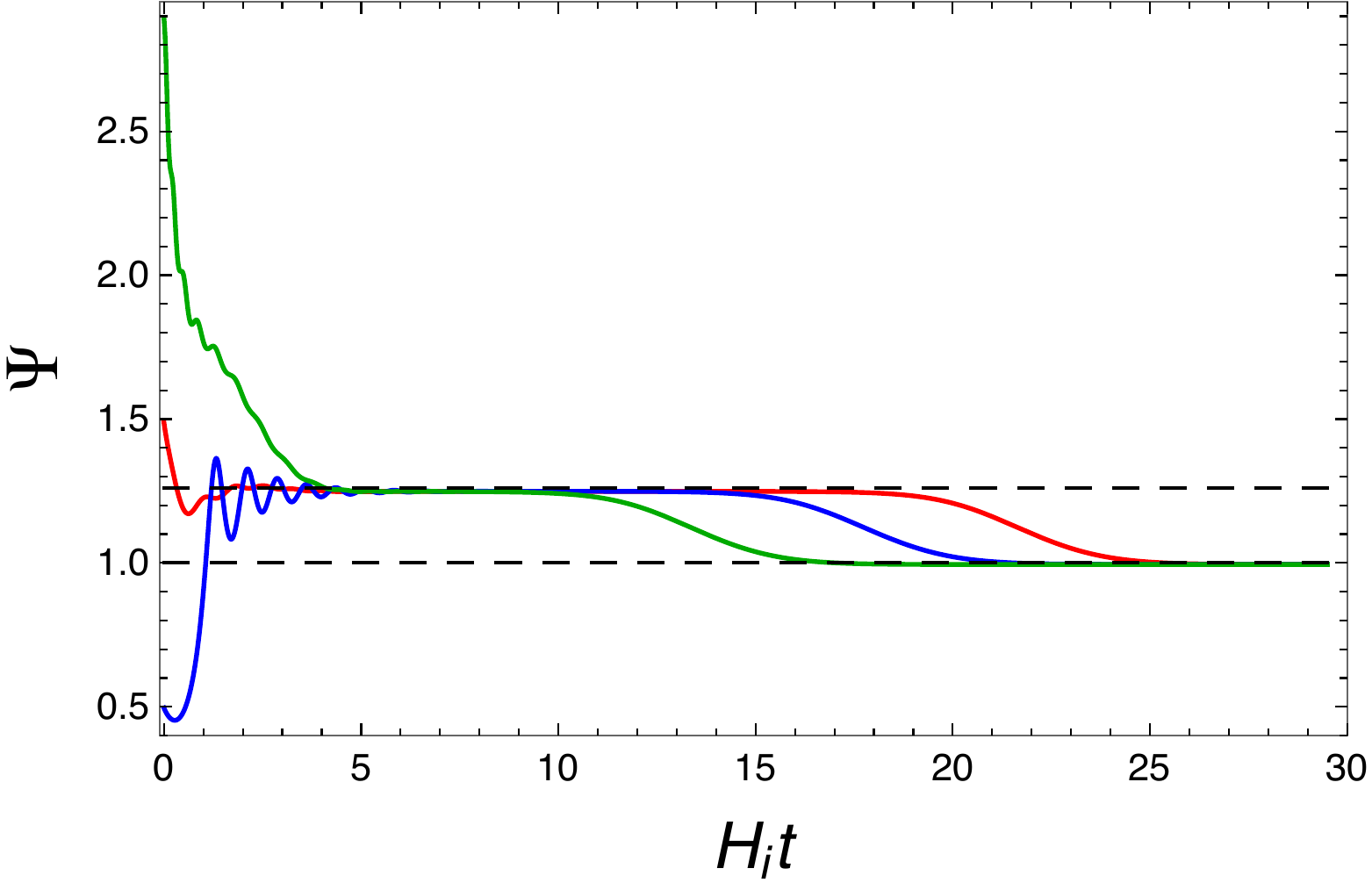}
        \includegraphics[keepaspectratio=true,height=37mm]{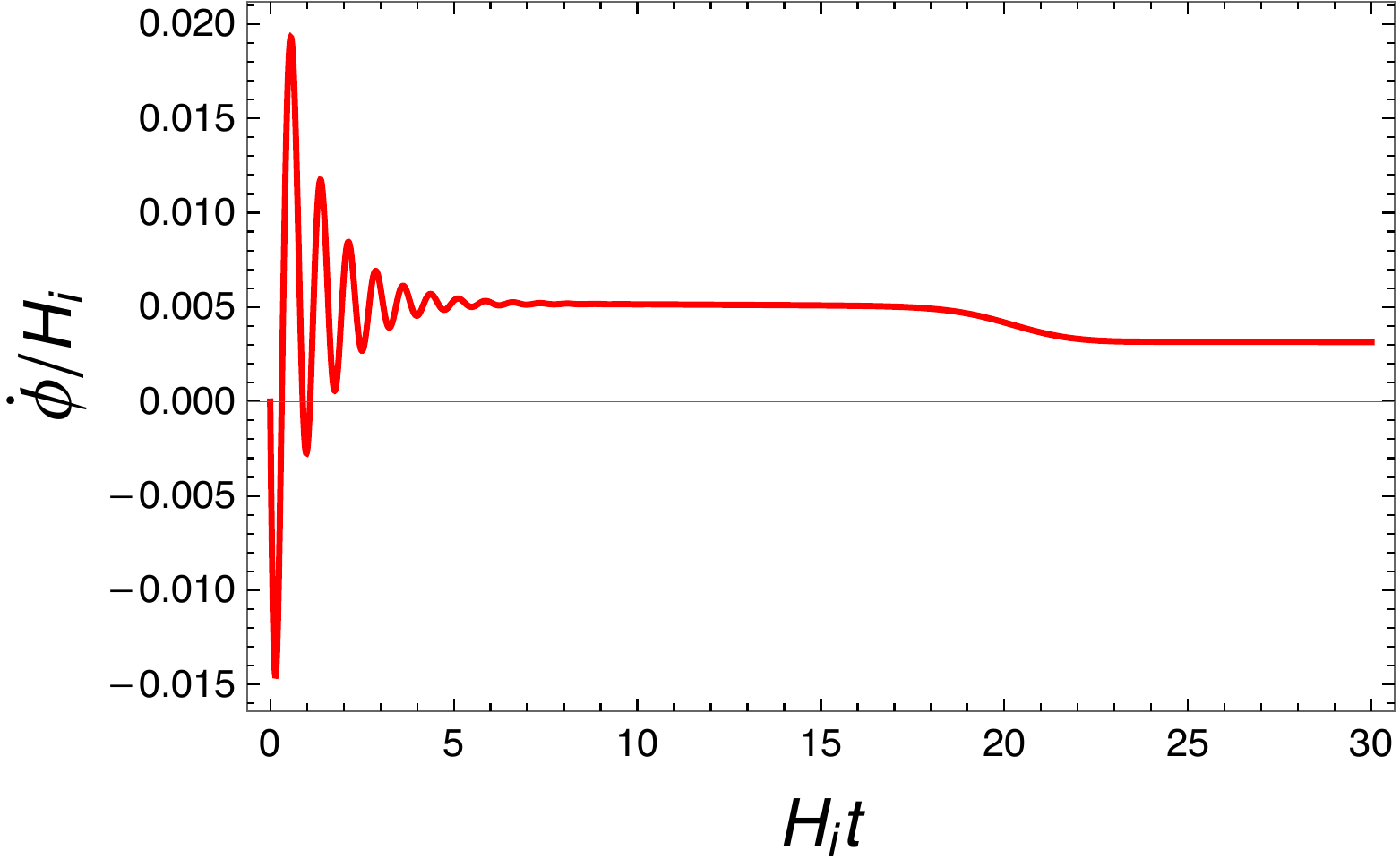}
        \includegraphics[keepaspectratio=true,height=37mm]{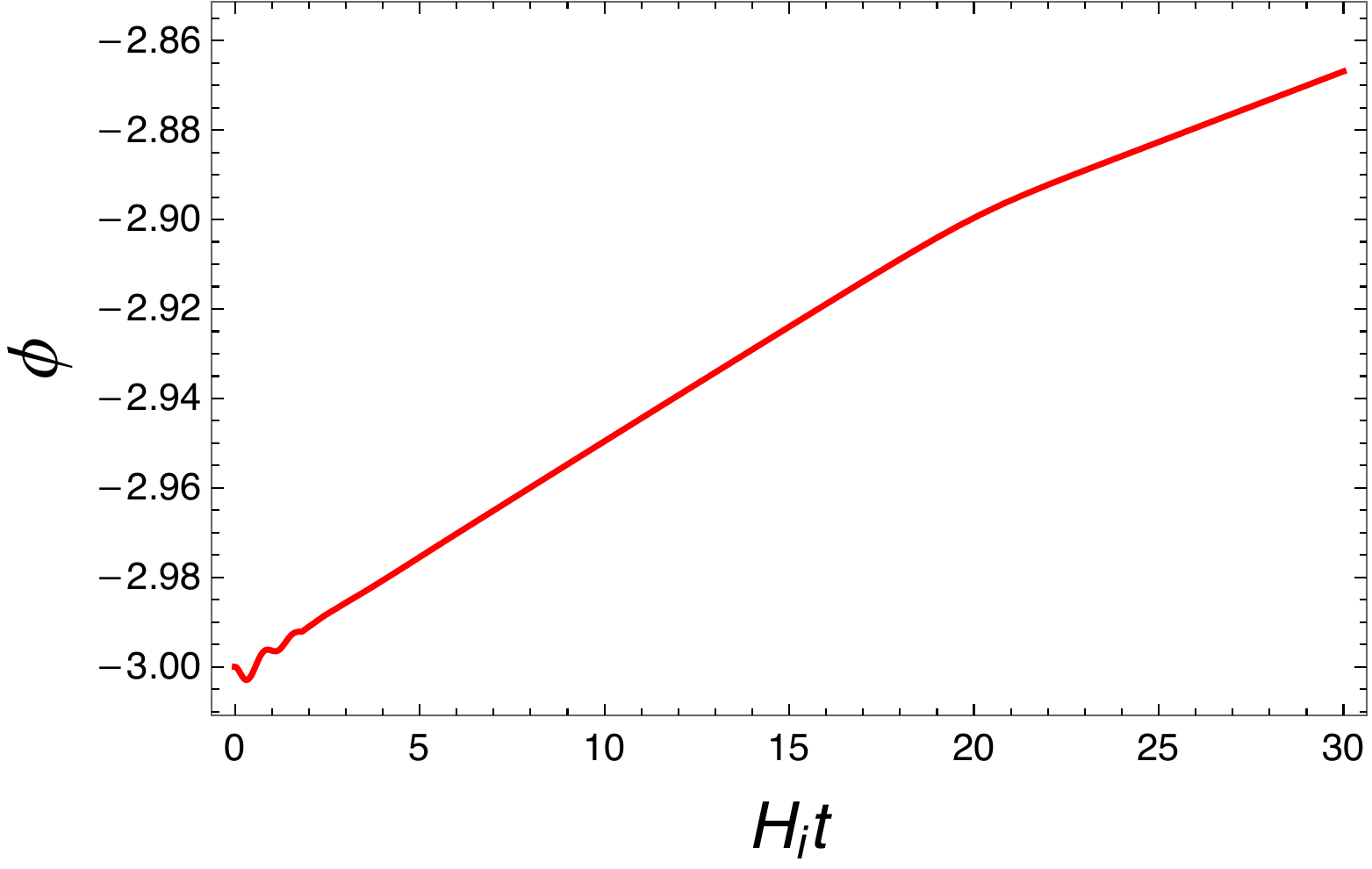}
    \end{center}
     \caption{The mean amplitude of the gauge field $\Psi(t)$ exhibiting the transitions from the $\lambda=1/2$ solution to the $\lambda=1$ one (left).
     For the blue and red lines, we impose the additional initial condition $\psi_5(0)=\psi_6(0)=0$ and employ a tiny but nonzero value of $\psi_8(0)$. 
     For the green line, we have nonzero initial values of $\psi_5(0),\psi_6(0)$, and $\psi_8(0)$, and they are comparable to the other components, i.e., the initial amplitudes are not small.
     However, the values of $\psi_4 \psi_6 + \psi_5 \psi_7$ and $\psi_8$ become very small for $5\lesssim H_i t\lesssim 10$.
     The black dashed horizontal lines are the same as those in Fig.~\ref{fig:plot32num_u1_compare_init.pdf}.
     We also plot the time evolution of $\dot\phi$ (center) and $\phi$ (right).
     The initial condition is the same as that used for plotting the red line in the left panel.
 	}
      \label{fig:plot_u1_transition.pdf}
 \end{figure}

The left panel of \figref{fig:plot_u1_transition.pdf} shows an example of
the mean amplitude of the gauge field with a tiny initial value of $\psi_8$ (the blue and red lines).
The system temporarily approaches the $\lambda=1/2$ isotropic solution
with negligibly small $\psi_5$, $\psi_6$, and $\psi_8$, but eventually these components become nonnegligible, forcing the system to evolve into the $\lambda=1$ isotropic
solution. The system thus exhibits a transition from an approximate isotropic state with $\lambda=1/2$
to the $\lambda=1$ isotropic final state. The moment at which the transition occurs depends on
the initial conditions, and during the transition the anisotropies in
the metric and the gauge field temporarily grow.
In the central and right panels in \figref{fig:plot_u1_transition.pdf}, we show the time evolution of $\dot\phi$ and $\phi$ at around the transition, respectively.
The evolution is not trivial (in the sense that they do not stay the same), but can be understood simply as follows. 
We can see that $\dot\phi$ becomes smaller after the transition, and this is because $\dot\phi$ is proportional to $\lambda^{2/3}$. 
The evolution of $\phi$ tracks that of $H$, because in the slow-roll regime $H$ is determined by the potential, which in the present model is equal to $\phi$ multiplied by the constant $-\mu^4 /f$.
In the above discussion, we considered the case with a tiny initial value of $\psi_8$.
However, we can develop the following comprehensive argument including tiny initial values of $\psi _5$ and $\psi _6$ altogether.

Now, we use linear analysis to investigate that the solution of $\lambda = 1/2$ can be
unstable and a transition can occur in the axisymmetric case.
We consider the $\lambda = 1/2$ isotropic solution where $\psi_1,\psi_4,\psi_7$ form the SU(2) subalgebra. Then, at leading order in the slow-roll approximation, $\psi_1,\psi_4,\psi_7$, and $\dot{\phi}$ are given in the form
\begin{align}
  &\psi_1 = \frac{2H}{\ga} m_{\psi},
  \quad
  \psi_4 = \psi_7 = \frac{\sqrt{2}H}{\ga} m_{\psi},
  \\
  &\frac{\dot{\phi}}{2fH} = m_{\psi} + \frac{1}{m_{\psi}},
  \label{eq:static_BG}
\end{align}
where $m_{\psi}$ is a dimensionless gauge field amplitude and
the amplitude of the gauge field is chosen to satisfy $\psi_1^2=\psi_4^2+\psi_7^2$.
Equation~\eqref{eq:static_BG} is determined from the background of Eq.~\eqref{eq:A1_u1_k=2} with $\ddot{\psi}_1,\dot{\psi}_1,\dot\sigma,\dot{H}=0$ and $\psi_1^2=\psi_4^2+\psi_7^2$.
On the top of the above background solution,
we introduce nonzero $\psi_5,\psi_6$, and $\psi_8$ as small perturbations.
The linearized equations of motion for $\psi_5,\psi_6$, and $\psi_8$ are written as
\begin{align}
  \mathbf{V}'' + 3\mathbf{V}' + {\cal M} \mathbf{V} = 0,
  \quad
  \mathbf{V} =
  \begin{pmatrix}
    \psi_5 \\
    \psi_6 \\
    \psi_8 \\
  \end{pmatrix}
  ,\quad
  {\cal M} =
  \begin{pmatrix}
    3 m_{\psi}^2 & 3 m_{\psi}^2 & -\sqrt{6} \\
    3 m_{\psi}^2 & 3 m_{\psi}^2 & -\sqrt{6} \\
    -2 \sqrt{6} & -2 \sqrt{6} & 2+6 m_{\psi}^2
  \end{pmatrix},
\end{align}
where a prime ($'$) denotes a derivative with respect to the e-folding number
$N:=\ln a$.
The matrix ${\cal M}$ can be diagonalized by some matrix $P$, and three eigenvalues are found to be $0,\, 6m_{\psi}^2+6$, and $6m_{\psi}^2 -4$.
The third eigenvalue can be negative when the normalized gauge field amplitude $m_{\psi}$ satisfies
\begin{align}
  -\sqrt{\frac{2}{3}} < m_{\psi} < \sqrt{\frac{2}{3}}.
  \label{con:param_of_GFA}
\end{align}
Then, the equation of motion for a certain linear combination of $\psi_5,\psi_6$, and $\psi_8$ associated to this negative eigenvalue reads
\begin{align}
  \tilde{v}'' +3\tilde{v}' 
  -\LL(4 - 6m_{\psi}^2\RR)\tilde{v}=0,
\end{align}
where $\tilde{v}$ is the third component of the vector $P^{-1}\mathbf{V}$.
The general solution of the above equation is
\begin{align}
  \tilde{v} 
  = c_1 a^{-\tfrac{3}{2}
  +\tfrac{1}{2}\sqrt{25-24 m_{\psi}^2}}
  + c_2 a^{-\tfrac{3}{2}
  -\tfrac{1}{2}\sqrt{25-24 m_{\psi}^2}},
  \label{eq:growing_mode}
\end{align}
where $c_1$ and $c_2$ are integration constants.
The first term grows under the condition \eqref{con:param_of_GFA}, and hence the $\lambda=1/2$ solution is unstable.

\begin{figure}[tb]
    \begin{center}
        \includegraphics[keepaspectratio=true,height=58mm]{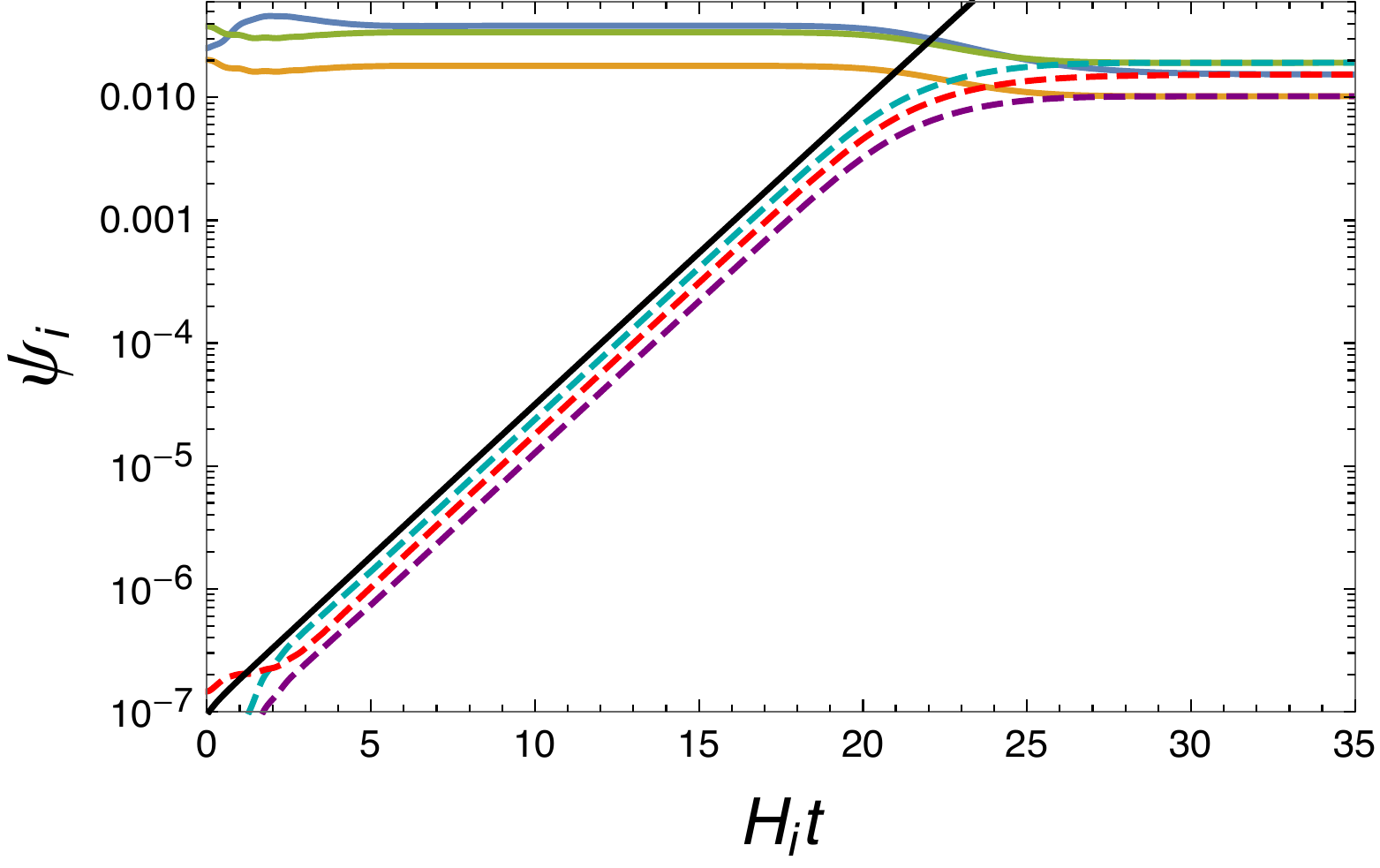}
    \end{center}
     \caption{
     The evolution of the gauge field amplitudes as compared to the analytic estimate of the growth rate.
     The background value of the gauge field are $\psi_1$ (blue), $\psi_4$ (orange) and
     $\psi_7$ (green).
     The perturbations of the gauge field are $\psi_5$ (dashed cyan), $\psi_6$ (dashed purple) and $\psi_8/\sqrt{3}$ (dashed red).
     The initial conditions of the background gauge field $\psi_1,\psi_4,\psi_7$ are a few times $0.01$, and for the perturbations of the gauge field we set the initial conditions $\psi_5=\psi_6=0$ and $\psi_8\sim 10^{-7}$.
     The black line shows the analytic estimate of the growth rate in Eq.~\eqref{eq:growing_mode}.
     One can see that the small perturbations grow and cause the transition.}
      \label{fig:plot42num_tr_compare_analytic.pdf}
 \end{figure}

Figure~\ref{fig:plot42num_tr_compare_analytic.pdf} shows an example of our numerical results for the evolution of the  gauge field components with tiny $\psi_8$ and $\psi_5=\psi_6=0$ at the initial time.
In this case, when the background components settle at the $\lambda=1/2$ isotropic solution, 
we have $\psi_1 \simeq 0.038 \mpl$,  $H\simeq 5.73 \times 10^{-7}\mpl$, and
$g_A=2\times 10^{-6}$, which leads to $m_\psi\simeq 0.067$.
Using Eq.~\eqref{eq:growing_mode}, we present our analytic estimate of the growth rate of these perturbations as the black line in \figref{fig:plot42num_tr_compare_analytic.pdf}. It is clear that the analytically obtained slope matches the numerical result very well.
Therefore, we confirm that the $\lambda=1/2$ solution is unstable
for $-\sqrt{2/3} < m_{\psi} < \sqrt{2/3}$
and it exhibits a transition to the $\lambda=1$ solution.
It is interesting to note that although the transient $\lambda=1/2$ solution
and the final $\lambda=1$ solution are both isotropic, a small anisotropy appears during the transition.

\begin{figure}[tb]
    \begin{center}
        \includegraphics[keepaspectratio=true,height=58mm]{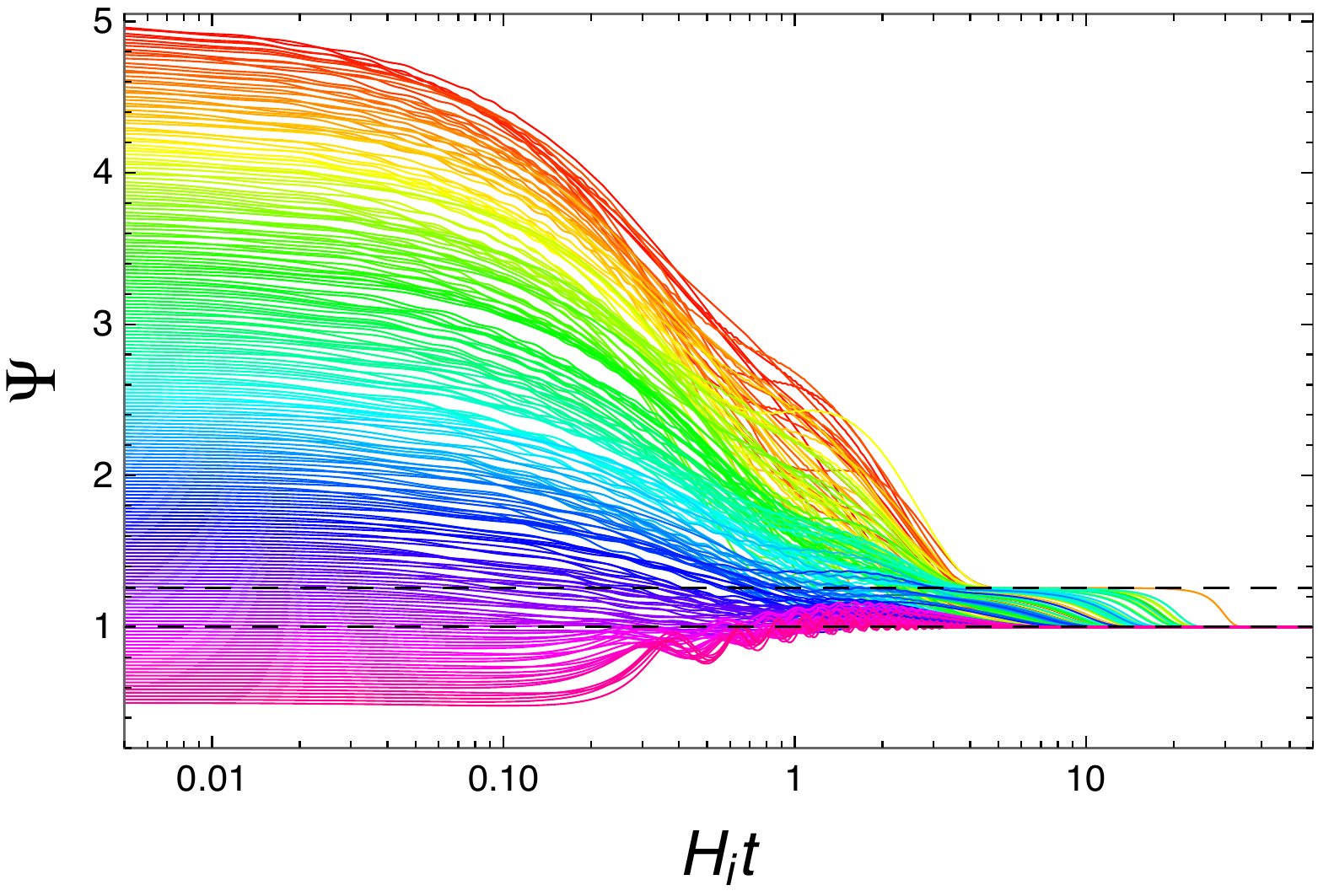}
    \end{center}
     \caption{
     The evolution of the gauge-field mean amplitude $\Psi(t)$.
     The initial conditions are chosen in the range $\Psi(0) \in [0.5,5]$ with positive random amplitudes of the gauge field.
     The parameter $\ga$ is set as $\ga =7\times 10^{-5}$.
     The colors show the initial value of $\Psi$.
     The black horizontal lines represent the isotropic solution~\eqref{eq:iso_const}, with the upper and lower lines corresponding to $\lambda=1/2$ and $\lambda=1$, respectively.
 	}
      \label{fig:plot33num_u1_random_init.pdf}
 \end{figure}

In \figref{fig:plot33num_u1_random_init.pdf}, we display 200 realizations of the mean amplitude of the gauge field~\eqref{eq:mean_amplitude} generated in the same way as the red lines in \figref{fig:plot32num_u1_compare_init.pdf}.
Here, we use the larger value for the coupling parameter, $\ga=7\times 10^{-5}$, to make $m_\psi$
a marginal value satisfying Eq.~\eqref{con:param_of_GFA}, that is, $m_{\psi}$ is slightly smaller than $\sqrt{2/3}$.
Then, the perturbations around the $\lambda=1/2$ solution have a small growth rate, and the gauge fields are expected to stay there 
for a longer time compared to the red line in \figref{fig:plot32num_u1_compare_init.pdf}.
Indeed, the transient stay at the $\lambda =1/2$ solution and the subsequent transition to the $\lambda =1$ solution are prominent in \figref{fig:plot33num_u1_random_init.pdf}.
We also expect that some solutions can be stabilized at the $\lambda =1/2$ solution if Eq.~\eqref{con:param_of_GFA} is violated, e.g. for an even larger coupling constant $\ga$.
Therefore, the evolution of the gauge fields depends on the value of $\ga$,
despite the universe being eventually isotropized in any case.

\begin{figure}[tb]
    \begin{center}
        \includegraphics[keepaspectratio=true,height=58mm]{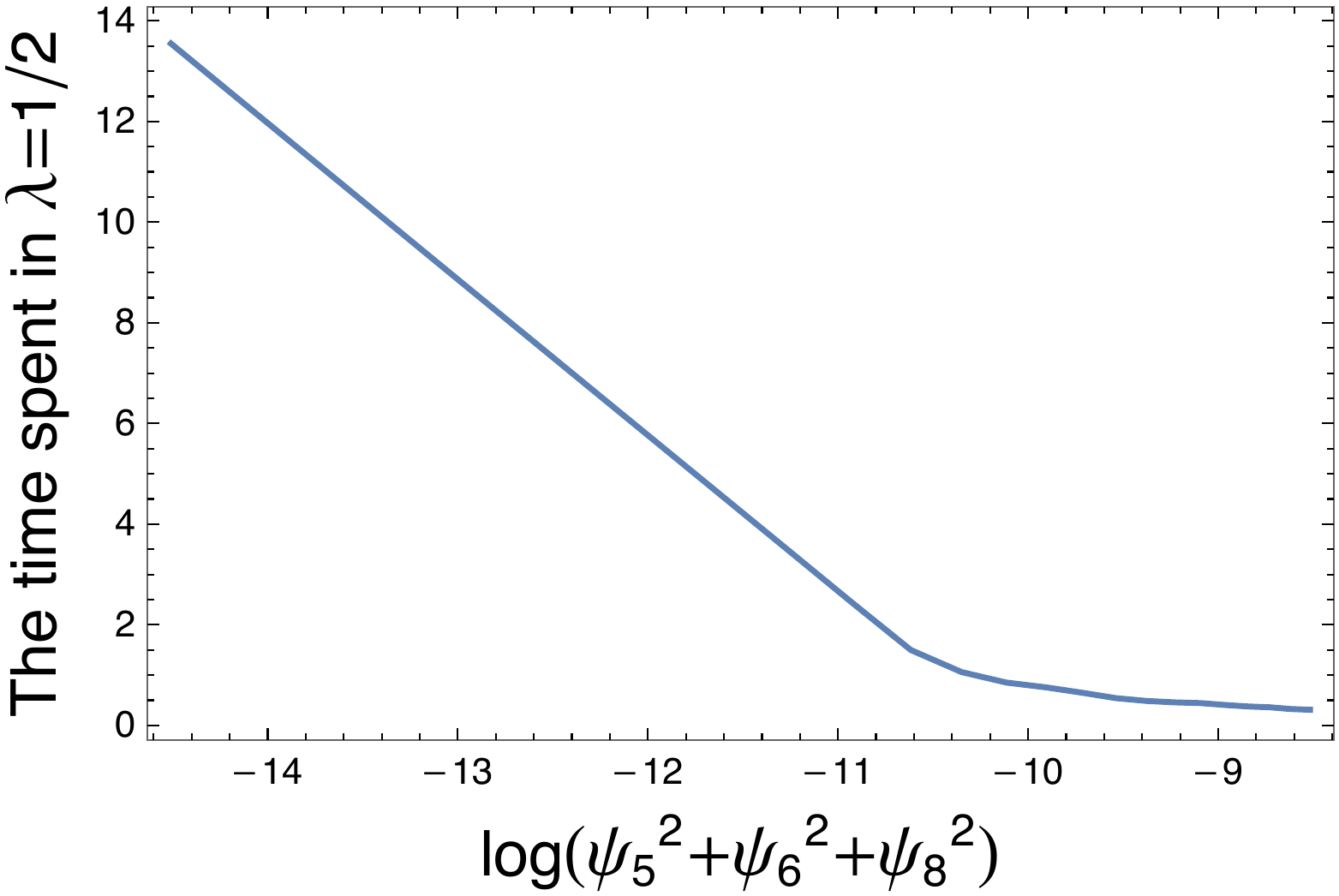}
    \end{center}
     \caption{
     The initial value of $\operatorname{log}( \psi_5^2 + \psi_6^2 + \psi_8^2)$ vs. the ``time spent close to $\lambda=1/2$'', which is defined as
     the duration that the solution is in the range $2^{1/3} \pm 0.01$. 
     The initial conditions for $\psi_1$, $\psi_4$, and $\psi_7$ are fixed
     as $(\psi_1,\psi_4,\psi_7)=(1.3\times 10^{-2}, 9.1\times 10^{-3}, 9.1\times 10^{-3})$, which leads to $\log(\psi_1^2+\psi_4^2+\psi_7^2)\approx -8$.
     }
      \label{fig:plot34num_u1_random_init_large_g_ref.pdf}
 \end{figure}

In \figref{fig:plot34num_u1_random_init_large_g_ref.pdf}, we show the initial value of $\operatorname{log}( \psi_5^2 + \psi_6^2 + \psi_8^2)$ v.s. the ``time spent close to $\lambda=1/2$'' with the initial conditions for $\psi_1$, $\psi_4$, and $\psi_7$ being fixed. Here, we defined the ``time spent close to $\lambda=1/2$'' as the duration that the solution is in the range $2^{1/3} \pm 0.01$. 
It can be seen that the smaller $\psi_5^2 + \psi_6^2 + \psi_8^2$ is, the longer the solution stays at $\lambda =1/2$.

\section{Conclusions and outlook}\label{sec:conclusions}

In this paper, we have studied the homogeneous
and anisotropic dynamics of SU($N$)-natural inflation.
We have focused mainly on axisymmetric gauge-field configurations
in the axisymmetric Bianchi type-I geometry to see whether and
how the anisotropies dilute in SU($N$)-natural inflation.
This simplified axisymmetric setup allows us to solve
the Einstein equations and the equations of motion for
the inflaton and gauge fields consistently
instead of relying on the test field approximation
in the fixed cosmological background,
thus improving the earlier analysis of Ref.~\cite{Fujita:2021eue}.

We have derived the equations to obtain axisymmetric configurations of
the SU($N$) gauge field and discussed the conditions under which
they are isotropic, and thus identified the candidates of
their late-time attractor.
On the basis of this argument, we have presented
an interesting explicit example
of axisymmetric configurations of the SU(3) gauge field
admitting two different SU(2) subalgebras characterized by
a number $\lambda$ appearing in the commutation relation
of the SU(2) generators. (In this case, we have $\lambda=1$ and $\lambda=1/2$.)
We have performed numerical calculations to see
the dynamics of this system in detail.
First, we have confirmed that SU($N$) natural inflation
does indeed isotropize the universe.
Second, and more interestingly, we have found that
under certain conditions the isotropic solution with
the $\lambda=1/2$ SU(2) subalgebra exhibits a transition to
the isotropic $\lambda=1$ solution which is therefore more stable.
This feature was not seen in the numerical study of the previous work~\cite{Fujita:2021eue}, essentially because of a simplifying assumption $\dot{\phi}=\,$const there. By performing a linear stability analysis, we found the condition for the transition~\eqref{con:param_of_GFA} and the growth rate of the instability~\eqref{eq:growing_mode}.

The study of cosmological perturbations in SU($N$)-natural inflation
has shown that the linear predictions depend on $\lambda$
through $\lambda\ga$, where $\ga$ is the gauge coupling constant,
implying that one cannot tell which isotropic solution is realized
at the level of linear perturbations~\cite{Fujita:2022fff}.
This degeneracy, however, is broken if the transition occurs
and $\lambda$ changes its value
during the observable period of inflation.
Since the amplitude of gravitational waves sourced by a SU($N$) gauge field is exponentially sensitive to the background gauge field amplitude~\cite{Fujita:2022fff} and the different isotropic solutions, such as $\lambda=1$ and $\lambda=1/2$ in the SU(3) case, have different amplitudes, the transition of the background solutions would lead to a step-like feature in  the spectrum of the sourced gravitational waves.
Therefore, the transition could leave a detectable signature on the B-mode polarization of the cosmic microwave background radiation or the stochastic gravitational wave background at the interferometer frequency range depending on when it occurs. It would be interesting to further study these potential observable imprints of SU($N$)-natural inflation and we leave it for future work.

\acknowledgments
The work of T. M. was supported by the Rikkyo University Special Fund for Research.
The work of T. F. was supported by JSPS KAKENHI Grants No.~JP18K13537 (T.F.) and No.~JP20H05854. The work of T. K. was supported by
JSPS KAKENHI Grant No.~JP20K03936 and
MEXT-JSPS Grants-in-Aid for Transformative Research Areas (A) ``Extreme universe,''
No.~JP21H05182 and No.~JP21H05189.

\appendix

\section{A QUICK RECAP OF SU(3)}\label{app:su3}

In this appendix, we very briefly summarize the SU(3) algebra.
The generators $T_a$ of SU($3$) can be chosen as
$T_a = \lambda _a /2$ with the Gell-Mann matrices
\begin{align}
  &\lambda_1 =
  \begin{pmatrix}
    0 & 1 & 0 \\
    1 & 0 & 0 \\
    0 & 0 & 0
  \end{pmatrix},
  \quad
  \lambda_2 =
  \begin{pmatrix}
    0 & -i & 0 \\
    i & 0 & 0 \\
    0 & 0 & 0
  \end{pmatrix},
  \quad
  \lambda_3 =
  \begin{pmatrix}
    1 & 0 & 0 \\
    0 & -1 & 0 \\
    0 & 0 & 0
  \end{pmatrix},
  \quad
  \lambda_4 =
  \begin{pmatrix}
    0 & 0 & 1 \\
    0 & 0 & 0 \\
    1 & 0 & 0
  \end{pmatrix},
  \\
  &\lambda_5 =
  \begin{pmatrix}
    0 & 0 & -i \\
    0 & 0 & 0 \\
    i & 0 & 0
  \end{pmatrix},
  \quad
  \lambda_6 =
  \begin{pmatrix}
    0 & 0 & 0 \\
    0 & 0 & 1 \\
    0 & 1 & 0
  \end{pmatrix},
  \quad
  \lambda_7 =
  \begin{pmatrix}
    0 & 0 & 0 \\
    0 & 0 & -i \\
    0 & i & 0
  \end{pmatrix},
  \quad
  \lambda_8 =
  \frac{1}{\sqrt{3}}
  \begin{pmatrix}
    1 & 0 & 0 \\
    0 & 1 & 0 \\
    0 & 0 & -2
  \end{pmatrix}.
\end{align}
These generators satisfy the commutation relation
\begin{align}
  \LL[T_a,T_b\RR] = if^{abc} T_c,
\end{align}
where $f^{abc}$ is the structure constant that is completely antisymmetric.
The structure constant is summarized in Table~\ref{tab:structure_SU(3)}.
\begin{table}[htb]
  \centering
    \caption{The nonvanishing structure constant of SU($3$).}
  \label{tab:structure_SU(3)}
  \begin{tabular}{|c||ccccccccc|} \hline
    $a b c$ & 123 & 147 & 156 & 246 & 257 & 345 & 367 & 458 & 678
    \\\hline
    $f^{abc}$ & ~~$1$~~ & ~$1/2$~ & $-1/2$ & ~$1/2$~ & ~$1/2$~ & ~$1/2$~ & $-1/2$ & $\sqrt{3} / 2$ & $\sqrt{3} / 2$
    \\\hline
  \end{tabular}
\end{table}

\section{THE CONDITIONS FOR ISOTROPIZATION IN GENERAL BIANCHI TYPE-I GEOMETRY}\label{app:c}

We extend the result of Sec.~\ref{Isotropization of metric} to the general Bianchi type-I universe.
The general Bianchi type-I metric is given by
\begin{align}
  \D s^2 = -\D t^2 + a^2(t) 
  \LL[
  e^{-4 \sigma_+(t)}\D x^2
  + e^{2\sigma_+(t) + 2\sqrt{3}\sigma_-(t)}\D y^2
  + e^{2\sigma_+(t) - 2\sqrt{3}\sigma_-(t)}\D z^2
  \RR].
\end{align}
In general Bianchi type-I universe, we decompose the SU($N$) gauge field in the temporal gauge $(A^a_0=0)$ as
\begin{align}
  A_i^a \D x^i
  = \psi_x^a (t) \LL[a(t) e^{-2\sigma_+ (t)} \RR] \D x
  + \psi_y^a (t) \LL[a(t) e^{\sigma_+(t) + \sqrt{3}\sigma_-(t)} \RR] \D y
  + \psi_z^a (t) \LL[a(t) e^{\sigma_+(t) - \sqrt{3}\sigma_-(t)} \RR] \D z,
\end{align}
where $i=1,2,3$ and $a = 1,\dots ,N^2-1$.
We define the norm of the gauge field for each $i$ as
\begin{align}
  \nmi(t)
  := \left[ \frac{1}{2}\sum_{a=1}^{N^2-1}\LL(\psi_i^a\RR)^2 \right]^{1/2}
  \quad(\text{no sum over } i).
\end{align}
This expression implies that $\sqrt{2}\nmi$ corresponds to the radial coordinate in the $\LL(N^2-1\RR)$-dimensional spherical coordinate system when $\psi_i^a$ is viewed as the Cartesian coordinate.
One can thus parametrize the components of
the SU($N$) gauge field
using the spherical coordinates for each $i$ as
\begin{align}
  \psi_i^a=\sqrt{2}\nmi(t)n_i^a(t),\qquad 
  n_i^a=\left\{
  \begin{array}{ll}
    \displaystyle
    \cos \theta^{a}_{i}
    \prod_{b=1}^{a-1} \sin \theta^{b}_{i} & (a=1, \dots, N^2-2)
    \\
    \displaystyle
    \prod_{b=1}^{N^2-2} \sin \theta^{b}_{i} & (a=N^2-1)
  \end{array}
  \right.,
  \qquad(\text{no sum over } i),
  \label{eq:polar_coord_comp1}
\end{align}
where $0\leq \theta^a_{i}(t) \leq \pi$ ($a=1,\dots,N^2-3$) and
$0\leq \theta^{N^2-2}_{i}(t) < 2\pi$.
Furthermore, it is convenient to decompose $\nmi$ into
the geometric mean and the anisotropic part as
\begin{align}
    \nmi=\nma(t)e^{\trf_i(t)},
    \quad 
    \nma:=\left[R_{x}R_{y}R_{z}\right]^{1/3},
    \quad 
    \sum_i^3 \trf_i=0.
\end{align}
From the symmetry we assume that $R_{y}=R_{z}$.
To summarize, we can express the component of the gauge field as
\begin{align}
  A^a_i = \psi^a_i a e^{\Sigma_i}
  = \sqrt{2} a\nma e^{\trf_i + \Sigma_i} n_i^a,
  \qquad(\text{no sum over } i),
\end{align}
where we introduce the vector as $\Sigma_i = \LL(-2\sigma_+ , \sigma_+ + \sqrt{3}\sigma_- , \sigma_+ - \sqrt{3}\sigma_- \RR)$
that represents the metric anisotropy.

The Chern-Simons term in the action can then be written as
\begin{align}
  F^a_{\mu\nu}\widetilde{F}_a^{\mu\nu}
  &= \frac{2}{3} \frac{\ga}{\sqrt{-g}}\partial_t
  \left(\epsilon^{ijk} f_{abc} A^a_i A^b_j A^c_k \right)
  \notag\\
  &=\frac{8\sqrt{2}\ga}{\sqrt{-g}}\partial_t
  \left[\LL(a\nma\RR)^3 f_{abc} n^a_x n^b_y n^c_z \right],
\end{align}
from which
we see that,
while $\nma(t)$ and the directions in the internal space, $n_i^a(t)$,
come into play,
the anisotropic part $\trf_i$
of the gauge field does not make any contributions to the Chern-Simons term.
The anisotropic part of the gauge field $\trf_i$ is not sourced by this term.
It implies that the inflaton's energy is not directly transferred to the anisotropic part and one naively expects that the anisotropic part simply decays as the universe expands in the same way as free gauge fields. Nevertheless, there is a possibility that the isotropic part of the gauge field sources it through self-interaction. To check this possibility we performed numerical calculations in the axisymmetric cases. As shown in the bulk of this paper (see \figref{fig:plot12num_u1_k=1_ds.pdf}), our numerical results indicate the anisotropic part is not sourced but quickly decays in the axisymmetric cases.
Therefore, we can also conjecture that the norm of the gauge field is isotropized $R_{i}\to \bar{R}$ in the general Bianchi type-I geometry.

We now extract the terms involving the metric anisotropy $\sigma_{\pm}$
from our Lagrangian to get
\begin{align}
  {\cal L}_{\sigma_+} 
  &= 3\mpl^2 \LL(\dot\sigma_+^2 + \dot\sigma_-^2\RR)
  - \dot\sigma_+ 
  \partial_t \LL(2R_{x}^2 - R_{y}^2 - R_{z}^2\RR)
  + \sqrt{3} \dot\sigma_- 
  \partial_t \LL(R_{y}^2 - R_{z}^2\RR)
  \notag\\
  &~~ -2H \dot\sigma_+ \LL(2R_{x}^2 - R_{y}^2 - R_{z}^2\RR)
  + 2\sqrt{3} \dot\sigma_- \LL(H + \dot\sigma_+\RR) \LL(R_{y}^2 - R_{z}^2\RR)
  + \dot\sigma_+^2 \LL(4R_{x}^2 + R_{y}^2 + R_{z}^2\RR)
  + 3\dot\sigma_-^2 \LL(R_{y}^2 + R_{z}^2\RR)
  .
\end{align}
Since there is a symmetry under a constant shift
of the metric anisotropy, $\sigma_{\pm}\to\sigma_{\pm}+\,$const,
we have a conserved quantity (denoted as ${\cal D_{\pm}}$).
One can thus integrate the equation of motion for $\sigma_{\pm}$
to obtain
\begin{align}
  &\dot\sigma_+ = \frac{{\cal D}_+/a^3 
  +\partial_t\LL(2R_{x}^2 - R_{y}^2 - R_{z}^2\RR)/2
  +H \LL(2R_{x}^2 -R_{y}^2 -R_{z}^2\RR)
  -\sqrt{3}\sigma_- \LL(R_{y}^2 - R_{z}^2\RR)}
  {3\mpl^2 + 4R_{x}^2 + R_{y}^2 + R_{z}^2},
  \\
  &\dot\sigma_- = \frac{{\cal D}_-/a^3 
  +\sqrt{3}\partial_t\LL(R_{y}^2 - R_{z}^2\RR)/2
  +\sqrt{3}\LL(H +\dot\sigma_+\RR) \LL(R_{y}^2 -R_{z}^2\RR)}
  {3\LL(\mpl^2 + R_{y}^2 + R_{z}^2\RR)},
\end{align}
If the anisotropic part of the gauge field decays away,
$\trf_{i}\to 0$, then one has $R_{i}\to \bar{R}$,
and hence $\dot\sigma_{\pm}\to 0$ at late times.
The spatial metric is therefore isotropized
if the norm of the gauge field defined by Eq.~\eqref{def:norm-a}
settles down to $R_{x}=R_{y}=R_{z}$ regardless of $n_i^a$.

\section{THE FIELD EQUATIONS}\label{sec:eoms}

In this appendix, we present the field equations
for the inflaton, the gauge field, and the metric.
As discussed in Sec.~\ref{sec:axisymmetric_SU(N)},
we investigate the gauge-field configurations characterized by
the conditions $u_1\neq 0$ and others$\,=0$,
and we have two cases: $u_1=1$ and $u_1=2$.

\subsection{Field equations for $u_1=1$}\label{sec:eom_u1=1}
The scalar-field and gauge-field equations of motion are given by
\begin{align}
  &\ddot\phi +3H\dot\phi -\frac{\mu^4}{f}\sin{\frac{\phi}{f}}
  = 
  -\frac{\ga\lambda}{f}
  \LL[
  \LL(\dot{\psi}_1 + 3H\psi_1\RR)
  \LL(\psi_2^2 + \psi_3^2\RR)
  + 2\psi_1\LL(
  \dot{\psi}_2 \psi_2
  + \dot\psi_3 \psi_3
  \RR)
  \RR],
  \label{eq:phi_u1_k=1}
  \\
  &\ddot{\psi}_1+3H\dot{\psi}_1
  +\psi_1\left[2\left(H+\dot{\sigma}\right)
  \left(H-2\dot{\sigma}\right) +\dot{H}-2\ddot{\sigma}
  +2\ga^2\LL(\psi_2^2 + \psi_3^2\RR)\right]
  =\frac{\ga\lambda}{f}\dot{\phi}\LL(\psi_2^2 + \psi_3^2\RR),
  \\
  &\ddot{\psi}_2+3H\dot{\psi}_2
  +\psi_2\left[\left( H+\dot{\sigma} \right) \left( 2H-\dot{\sigma} \right) +\dot{H}+\ddot{\sigma}
  +\ga^2\left(\psi_1^2+\psi_2^2 +\psi_3^2 \right)\right]
  =\frac{\ga\lambda}{f}\dot{\phi}\psi_1\psi_2,
  \\
  &\ddot{\psi}_3+3H\dot{\psi}_3
  +\psi_3\left[\left( H+\dot{\sigma} \right) \left( 2H-\dot{\sigma} \right) +\dot{H}+\ddot{\sigma}
  +\ga^2\left(\psi_1^2+\psi_2^2 +\psi_3^2 \right)\right]
  =\frac{\ga\lambda}{f}\dot{\phi}\psi_1\psi_3,
  \\
  &\ddot{\psi}_8 + 3H\dot{\psi}_8
  +\psi_8\LL[2\LL(H+\dot\sigma\RR)\LL(H-2\dot\sigma\RR)
  +\dot{H} -2\ddot\sigma\RR]=0.
  \label{psi8 EoM u1=1 case}
\end{align}
The Einstein equations read
\begin{align}
  \mpl^2 \LL(2\dot{H} + 3H^2 + 3\dot\sigma^2\RR) 
  =&~ -p_{\phi} -p_{1},
  \label{eq:accel_u1_k=1}\\
  3\mpl^2 \LL(\ddot{\sigma} + 3H\dot{\sigma}\RR)
  =&
  \sum_{a=1,8} \left[\dot{\psi}_a +\psi_a \left(H -2\dot{\sigma}\right)\right]^2
  - \sum_{a=2,3} \left[\dot{\psi}_a +\psi_a \left(H +\dot{\sigma}\right)\right]^2
  \notag\\
  &- \ga^2 \LL(\psi_1^2 -\psi_2^2 -\psi_3^2 \RR)\LL(\psi_2^2 +\psi_3^2 \RR),
  \label{eq:sigma_u1_k=1}\\
  3\mpl^2 \LL(H^2 - \dot\sigma^2\RR) 
  =&~ \rho_{\phi} + \rho_{1},
  \label{eq:Friedmann_u1=1}
\end{align}
where
\begin{align}
  \rho_{\phi} =&~ \frac{1}{2}\dot\phi^2 - \frac{\mu^4}{f}\phi,
  \\
  p_{\phi} =&~ \frac{1}{2}\dot\phi^2 + \frac{\mu^4}{f}\phi,
  \\
  \rho_{1} =& ~
  \frac{1}{2}\sum_{a=1,8} \left[\dot{\psi}_a +\psi_a \left(H -2\dot{\sigma}\right)\right]^2
  + \sum_{a=2,3} \left[\dot{\psi}_a +\psi_a \left(H +\dot{\sigma}\right)\right]^2
  +\frac{\ga^2}{2} \left(\psi _2^2+\psi _3^2\right)
  \left(2 \psi _1^2+\psi _2^2+\psi _3^2\right),
  \label{eq:rho_u_1=1}
  \\
  p_{1} =&~ \frac{\rho_{1}}{3}.
\end{align}
The equation of motion for the metric anisotropy
can be integrated once to give
\begin{align}
  \dot{\sigma} = \frac{{\cal D}/a^3+\dot{\psi}_1\psi_1 -\dot{\psi}_2\psi_2 -\dot{\psi}_3\psi_3 +\dot{\psi}_8\psi_8 
  +H\LL(\psi_1^2 -\psi_2^2 -\psi_3^2 +\psi_8^2\RR)}
  {3 \mpl^2+2 \psi_1^2+\psi_2^2+\psi_3^2+2 \psi_8^2},
  \label{eq:metric_aniso_u1=1}
\end{align}
where ${\cal D}$ is an integration constant.
We see that if
\begin{align}
  \psi_1^2(t) +\psi_8^2(t) = \psi_2^2(t) +\psi_3^2(t),
\end{align}
then $\dot\sigma\to 0$ at late times (${\cal D}/a^3\to 0$)
and we have the isotropic universe.
In this configuration, the gauge constraint is obtained as 
\begin{align}
  &e^{4\sigma} a^4 \LL(\psi_2\dot{\psi}_3 - \psi_3\dot{\psi}_2 \RR) = 0.
  \label{eq:gauge_constraint_u1=1}
\end{align}
Therefore, the gauge constraint is automatically satisfied if 
$\dot{\psi}_2=\dot{\psi}_3=0$ holds.

\subsection{Field equations for $u_1=2$}\label{sec:eom_u1=2}
We write the gauge field as
\begin{align}
  A_x^1 = ae^{-2\sigma}\psi_1, \quad
  A_y^4 = ae^{\sigma}\psi_4, \quad
  A_y^5 = ae^{\sigma}\psi_5, \quad
  A_y^6 = ae^{\sigma}\psi_6, \quad
  A_y^7 = ae^{\sigma}\psi_7, \quad
  A_x^8 = ae^{-2\sigma}\psi_8.
  \label{eq:A_to_psi_k=2}
\end{align}
Then, the scalar-field and gauge-field equations of motion are
\begin{align}
  &\ddot\phi +3H\dot\phi -\frac{\mu^4}{f}\sin{\frac{\phi}{f}}
  =-\frac{\ga\lambda}{2f}\sum_{a=4}^{7}
  \psi_1\psi_a^2
  \LL[3H + \LL(\frac{\dot{\psi}_1}{\psi_1} +2\frac{\dot{\psi}_a}{\psi_a} \RR)\RR]
  \notag\\
  & ~~ -\frac{\sqrt{3}\ga\lambda}{f}\psi_8
  \LL[
  \LL(\psi_4\psi_6 + \psi_5\psi_7\RR)
  \LL(3H
  + \frac{\dot\psi_8}{\psi_8}\RR)
  + \psi_4\psi_6
  \LL( \frac{\dot\psi_4}{\psi_4}
  + \frac{\dot\psi_6}{\psi_6}\RR)
  +\psi_5\psi_7
  \LL(
  \frac{\dot\psi_5}{\psi_5}
  + \frac{\dot\psi_7}{\psi_7}\RR)\RR],
  \label{eq:phi_u1_k=2}
  \\
  &\ddot{\psi}_1 + 3H\dot{\psi}_1
  +\psi_1\LL[2\LL(H+\dot\sigma\RR)\LL(H-2\dot\sigma\RR)
  +\dot{H} -2\ddot\sigma\RR]
  \notag\\
  & ~~~ +\frac{\ga^2}{2}\LL[\psi_1\LL(\psi_4^2 + \psi_5^2
  +\psi_6^2 +\psi_7^2\RR)
  + 2\sqrt{3}\psi_8\LL(\psi_4\psi_6 +\psi_5\psi_7\RR)\RR]
  =\frac{\ga\lambda}{2f}\dot\phi
  \LL(\psi_4^2 +\psi_5^2+\psi_6^2 +\psi_7^2\RR),
  \label{eq:A1_u1_k=2}\\
  &\ddot{\psi}_4 + 3H\dot{\psi}_4
  +\psi_4\LL[\LL(H+\dot\sigma\RR)\LL(2H-\dot\sigma\RR)
  +\dot{H} +\ddot\sigma\RR]
  \notag\\
  & ~~~ +\frac{\ga^2}{4}\LL[\psi_4\LL(\psi_1^2 + \psi_4^2 + \psi_5^2 +7\psi_6^2 +\psi_7^2 +3\psi_8^2\RR)
  + 2\psi_6\LL(3\psi_5\psi_7 + \sqrt{3}\psi_1\psi_8\RR)\RR]
  = \frac{\ga\lambda}{2f}\dot\phi
  \LL(\psi_1\psi_4 +\sqrt{3}\psi_6\psi_8\RR),
  \label{eq:A4_u1_k=2}\\
  &\ddot{\psi}_5 + 3H\dot{\psi}_5
  +\psi_5\LL[\LL(H+\dot\sigma\RR)\LL(2H-\dot\sigma\RR)
  +\dot{H} +\ddot\sigma\RR]
  \notag\\
  & ~~~ +\frac{\ga^2}{4}\LL[\psi_5\LL(\psi_1^2 + \psi_4^2 + \psi_5^2 +\psi_6^2 +7\psi_7^2 +3\psi_8^2\RR)
  + 2\psi_7\LL(3\psi_4\psi_6 + \sqrt{3}\psi_1\psi_8\RR)\RR]
  = \frac{\ga\lambda}{2f}\dot\phi
  \LL(\psi_1\psi_5 +\sqrt{3}\psi_7\psi_8\RR),
  \label{eq:A5_u1_k=2}\\
  &\ddot{\psi}_6 + 3H\dot{\psi}_6
  +\psi_6\LL[\LL(H+\dot\sigma\RR)\LL(2H-\dot\sigma\RR)
  +\dot{H} +\ddot\sigma\RR]
  \notag\\
  & ~~~ +\frac{\ga^2}{4}\LL[\psi_6\LL(\psi_1^2 + 7\psi_4^2 + \psi_5^2 +\psi_6^2 +\psi_7^2 +3\psi_8^2\RR)
  + 2\psi_4\LL(3\psi_5\psi_7 + \sqrt{3}\psi_1\psi_8\RR)\RR]
  = \frac{\ga\lambda}{2f}\dot\phi
  \LL(\psi_1\psi_6 +\sqrt{3}\psi_4\psi_8\RR),
  \label{eq:A6_u1_k=2}\\
  &\ddot{\psi}_7 + 3H\dot{\psi}_7
  +\psi_7\LL[\LL(H+\dot\sigma\RR)\LL(2H-\dot\sigma\RR)
  +\dot{H} +\ddot\sigma\RR]
  \notag\\
  & ~~~ +\frac{\ga^2}{4}\LL[\psi_7\LL(\psi_1^2 + \psi_4^2 + 7\psi_5^2 +\psi_6^2 +\psi_7^2 +3\psi_8^2\RR)
  + 2\psi_5\LL(3\psi_4\psi_6 + \sqrt{3}\psi_1\psi_8\RR)\RR]
  = \frac{\ga\lambda}{2f}\dot\phi
  \LL(\psi_1\psi_7 +\sqrt{3}\psi_5\psi_8\RR),
  \label{eq:A7_u1_k=2}\\
  &\ddot{\psi}_8 + 3H\dot{\psi}_8
  +\psi_8\LL[2\LL(H+\dot\sigma\RR)\LL(H-2\dot\sigma\RR)
  +\dot{H} -2\ddot\sigma\RR]
  \notag\\
  & ~~~ +\frac{\ga^2}{2}\LL[3\psi_8\LL(\psi_4^2 + \psi_5^2 +\psi_6^2 +\psi_7^2\RR) + 2\sqrt{3}\psi_1\LL(\psi_4\psi_6 +\psi_5\psi_7\RR)\RR]
  =\frac{\sqrt{3}\ga\lambda}{f}\dot\phi\LL(\psi_4\psi_6 +\psi_5\psi_7\RR).
  \label{eq:A8_u1_k=2}
\end{align}
The Einstein equations read
\begin{align}
  \mpl^2 \LL(2\dot{H} + 3H^2 + 3\dot\sigma^2\RR) =&~
  -p_{\phi} -p_{2},
  \label{eq:accel_u1_k=2}\\
  3\mpl^2 \LL(\ddot{\sigma} + 3H\dot{\sigma}\RR)
  =&
  \sum_{a=1,8} \left[\dot{\psi}_a +\psi_a \left(H -2\dot{\sigma}\right)\right]^2
  - \sum_{a=4}^{7} \left[\dot{\psi}_a +\psi_a \left(H +\dot{\sigma}\right)\right]^2
  \notag\\
  &
  -\frac{\ga^2}{4} \left(\psi _1^2+3 \psi _8^2\right)
  \left(\psi _4^2+\psi _5^2+\psi _6^2+\psi _7^2\right)
  +\frac{\ga^2}{4} \left(\psi _4^2 +\psi _7^2\right)^2
  +\frac{\ga^2}{4} \left(\psi _5^2 +\psi _6^2\right)^2
  \notag\\
  &
  +\ga^2 \left(\psi _4^2 \psi _6^2 + \psi _5^2 \psi _7^2\right)
  -\frac{\ga^2}{2} \left(\psi _4 \psi _5 + \psi _6 \psi _7\right)^2
  \notag\\
  &
  +\frac{5}{2}\ga^2 \left(\psi _4 \psi _6 + \psi _5 \psi _7\right)^2
  -\sqrt{3}\ga^2 \psi_1 \psi_8\left(\psi _4 \psi _6 + \psi _5 \psi _7\right),
  \label{eq:sigma_u1_k=2}\\
  3\mpl^2 \LL(H^2 - \dot\sigma^2\RR) =&~ \rho_{\phi} + \rho_{2},
\end{align}
where
\begin{align}
  \rho_{2} =& ~
  \frac{1}{2}\sum_{a=1,8} \left[\dot{\psi}_a +\psi_a\left(H -2\dot{\sigma}\right)\right]^2
  +\sum_{a=4}^{7} \left[\dot{\psi}_a +\psi_a \left(H +\dot{\sigma}\right)\right]^2
  \notag\\
  &
  +\frac{\ga^2}{4} \left(\psi _1^2+3 \psi _8^2\right)
  \left(\psi _4^2+\psi _5^2+\psi _6^2+\psi _7^2\right)
  +\frac{\ga^2}{8} \left(\psi _4^2 +\psi _7^2\right)^2
  +\frac{\ga^2}{8} \left(\psi _5^2 +\psi _6^2\right)^2
  \notag\\
  &
  -\frac{\ga^2}{4} \left(\psi _4 \psi _5 - \psi _6 \psi _7\right)^2
  +\frac{7}{4}\ga^2 \left(\psi _4 \psi _6 + \psi _5 \psi _7\right)^2
  +\sqrt{3}\ga^2 \psi_1 \psi_8\left(\psi _4 \psi _6 + \psi _5 \psi _7\right),
  \\
  p_{2} =&~ \frac{\rho_{2}}{3}.
\end{align}
The equation of motion for the metric anisotropy,
$\sigma$, can be integrated once to give
\begin{align}
  \dot{\sigma} = \frac{{\cal D}'/a^3+\dot{\psi_1}\psi_1 +\dot{\psi_8}\psi_8 -\dot{\psi_4}\psi_4 -\dot{\psi_5}\psi_5 -\dot{\psi_6}\psi_6 -\dot{\psi_7}\psi_7
  +H\LL(\psi_1^2 +\psi_8^2 -\psi_4^2 -\psi_5^2 -\psi_6^2 -\psi_7^2\RR)}
  {3 \mpl^2+2 \psi_1^2 +\psi_4^2 +\psi_5^2 +\psi_6^2 +\psi_7^2 +2\psi_8^2},
\end{align}
where ${\cal D}'$ is an integration constant.
We see that if
\begin{align}
  \psi_1^2(t) +\psi_8^2(t) = \psi_4^2(t) +\psi_5^2(t) +\psi_6^2(t) +\psi_7^2(t),
\end{align}
then we have $\dot\sigma\to 0$ at late times (${\cal D}'/a^3\to 0$)
and thus the universe becomes isotropic.
We need to consider gauge constraints, i.e.,
\begin{align}
  \partial_{i} F^{i 0}_a + \ga f^{abc} A_i^b F^{i0}_c = 0.
\end{align}
In our configuration, these components are obtained as
\begin{align}
  &e^{4\sigma} a^4 \LL(\psi_4\dot{\psi}_7 - \psi_7\dot{\psi}_4
  + \psi_6\dot{\psi}_5 - \psi_5\dot{\psi}_6\RR) = 0,
  \\
  &e^{4\sigma} a^4 \LL(\psi_4\dot{\psi}_5 - \psi_5\dot{\psi}_4
  + \psi_6\dot{\psi}_7 - \psi_7\dot{\psi}_6\RR) = 0.
\end{align}
Thus, the gauge constraints are automatically satisfied if the initial condition is taken as $\dot{\psi}_a=0$ for $a=4,\dots,7$.

\bibliography{refs}
\bibliographystyle{JHEP}

\end{document}